\documentclass[twocolumn, pdflatex]{aastex62}
\pdfoutput=1

\received{}
\revised{}
\accepted{}

\submitjournal{ApJ}

\shorttitle{Morphological independence on star formation efficiency}
\shortauthors{Koyama et al.}

\begin{document}
\title{Do Galaxy Morphologies Really Affect the Efficiency of Star Formation during the Phase of Galaxy Transition?}

\correspondingauthor{Shuhei Koyama}
\email{skoyama@cosmos.phys.sci.ehime-u.ac.jp}

\author[0000-0002-0100-1238]{Shuhei Koyama}
\affiliation{Research Center for Space and Cosmic Evolution, Ehime University, 2-5 Bunkyo-cho, Matsuyama, Ehime 790-8577, Japan}

\author{Yusei Koyama}
\affiliation{Subaru Telescope, National Astronomical Observatory of Japan, 650 North A'ohoku Place, Hilo, HI 96720, USA}
\affiliation{Department of Astronomical Science, Graduate University for Advanced Studies (SOKENDAI), 2-21-1 Osawa, Mitaka, Tokyo 181-8588, Japan}

\author{Takuji Yamashita}
\affiliation{Research Center for Space and Cosmic Evolution, Ehime University, 2-5 Bunkyo-cho, Matsuyama, Ehime 790-8577, Japan}

\author{Masao Hayashi}
\affiliation{National Astronomical Observatory of Japan, 2-21-1 Osawa, Mitaka, Tokyo 181-8588, Japan}

\author{Hideo Matsuhara}
\affiliation{Institute of Space and Astronautical Science, Japan Aerospace Exploration Agency, 3-1-1 Yoshinodai, Chuo-ku, Sagamihara, Kanagawa 252-5210, Japan}
\affiliation{Department of Space and Astronautical Science, Graduate University for Advanced Studies (SOKENDAI),
3-1-1 Yoshinodai, Chuo-ku, Sagamihara, Kanagawa 252-5210, Japan}

\author[0000-0002-6660-9375]{Takao Nakagawa}
\affiliation{Institute of Space and Astronautical Science, Japan Aerospace Exploration Agency, 3-1-1 Yoshinodai, Chuo-ku, Sagamihara, Kanagawa 252-5210, Japan}

\author{Shigeru V. Namiki}
\affiliation{Department of Astronomical Science, Graduate University for Advanced Studies (SOKENDAI), 2-21-1 Osawa, Mitaka, Tokyo 181-8588, Japan}

\author{Tomoko L. Suzuki}
\affiliation{National Astronomical Observatory of Japan, 2-21-1 Osawa, Mitaka, Tokyo 181-8588, Japan}
\affiliation{Astronomical Institute, Tohoku University, 63 Aramaki, Aoba-ku, Sendai 980-8578, Japan}

\author{Nao Fukagawa}
\affiliation{Department of Astronomical Science, Graduate University for Advanced Studies (SOKENDAI), 2-21-1 Osawa, Mitaka, Tokyo 181-8588, Japan}

\author{Tadayuki Kodama}
\affiliation{Astronomical Institute, Tohoku University, 63 Aramaki, Aoba-ku, Sendai 980-8578, Japan}

\author{Lihwai Lin}
\affiliation{Institute of Astronomy \& Astrophysics, Academia Sinica, Taipei 10617, Taiwan}

\author{Kana Morokuma-Matsui}
\affiliation{Institute of Space and Astronautical Science, Japan Aerospace Exploration Agency, 3-1-1 Yoshinodai, Chuo-ku, Sagamihara, Kanagawa 252-5210, Japan}

\author{Rhythm Shimakawa}
\affiliation{Subaru Telescope, National Astronomical Observatory of Japan, 650 North A'ohoku Place, Hilo, HI 96720, USA}

\author{Ichi Tanaka}
\affiliation{Subaru Telescope, National Astronomical Observatory of Japan, 650 North A'ohoku Place, Hilo, HI 96720, USA}

\begin{abstract}

\noindent
Recent simulations predict that the presence of stellar bulge suppress the efficiency of star formation in early-type galaxies, and this ``morphological quenching'' scenario is supported by many observations. In this study, we discuss the net effect of galaxy morphologies on the star formation efficiency (SFE) during the phase of galaxy transition, on the basis of our CO($J=1-0$) observations of 28 local ``green-valley'' galaxies with the Nobeyama 45m Radio Telescope. We observed 13 ``disk-dominated'' and 15 ``bulge-dominated'' green-valley galaxies at fixed stellar mass ($M_*$) and star formation rate (SFR), supplemented by 1 disk- and 6 bulge-dominated galaxies satisfying the same criteria from the xCOLD~GASS survey. By using a total of 35 green-valley galaxies, we reveal that the distributions of molecular gas mass, molecular gas fraction, and SFE of green-valley galaxies do {\it not} change with their morphologies, suggesting little impact of galaxy morphologies on their SFE, and interestingly this result is also valid for normal star-forming galaxies on the SF main-sequence selected from the xCOLD~GASS galaxies. On the other hand, we find that $\sim$20\% of bulge-dominated green-valley galaxies do not show significant CO emission line, showing {\it high} SFEs for their M$_*$ and SFR. These molecular gas deficient sources identified only in the bulge-dominated green-valley galaxies may represent an important population during the quenching phase under the influence of stellar bulge, but our results suggest that the presence of stellar bulge does not decrease the efficiency of on-going star formation, in contrast to the prediction of the morphological quenching scenario. 

\end{abstract}

%%%%%%%%%%  START: INTRODUCTION  %%%%%%%%%%
\section{introduction} \label{sec:intro}

Over the last decade, it has been established that star-forming galaxies (mostly with disk-dominated morphologies) exhibit a tight positive correlation between star formation rate (SFR) and stellar mass (M$_*$) across environment---i.e.\ so-called star-forming main sequence \citep[MS;][]{Dad07,Elb07,Noe07,Pen10,Koy13}, while red/passive galaxies (mostly with bulge-dominated morphologies) are distributed far below the MS by $\sim$1.5--2 dex on the SFR--M$_*$ diagram.
It is believed that galaxies located on the MS eventually evolve into the passive galaxy population by quenching their star formation, accompanying morphological transformation \citep[e.g.][]{Wuy11}.
The key questions are ``what" triggers this transition event and ``how" galaxies migrate from the MS to the passive galaxy sequence.

An interesting scenario proposed for the star formation (SF) quenching mechanism is the so-called ``morphological quenching" scenario \citep{Mar09}, where galaxy morphologies themselves control their star formation acitivity. \citet{Mar09} predicted that gas disks can be stabilized against star formation when they are embedded in a stellar spheroid, resulting in a lower SF efficiency in early-type galaxies.
A recent observational campaign of nearby early-type galaxies \citep[the ATLAS$^\mathrm{3D}$ project;][]{Cap11} and their molecular gas follow-up observations \citep[e.g.][]{You11, Mar13, Dav14} have shown that star-forming regions in early-type and late-type galaxies follow a different SF law \citep[Kennicutt--Schmidt law;][]{Sch59, Ken98} in the sense that star formation in early-type galaxies are less efficient by a factor of $\sim$2$\times$ compared to those in late-type galaxies, consistent with the morphological quenching scenario.
However, many of these earlier works attempted to compare typical early-type and late-type galaxies, and consequently ``early-type'' galaxies are usually selected from passive galaxy population, while ``late-type'' galaxies are selected from star-forming population in the MS.
A concern here is that the two populations are in very different evolutionary stages of galaxies, and that it may not be possible to assess the real effects of galaxy morphologies on the SF quenching. 
In other words, the claim that early-type galaxies have lower efficiency of star formation than late-type galaxies does not necessarily indicate that they are really quenched by the morphological quenching process.

A potential approach to directly assess the real impact of galaxy morphologies on the SF efficiency is to compare early-type and late-type galaxies in the same evolutionary stage. 
In this context, "green-valley galaxies" \citep{Sal14} are of great interest, because they are expected to be in the SF quenching phase \citep{Bel04, Fab07, Mar07}.
Green-valley galaxies are originally defined as an intermediate population located in between the "red sequence" and "blue cloud" in the color-magnitude diagram (CMD).
However, this definition may misidentify galaxies in the SF quenching, because the red colors of galaxies have a finite limit, which could lead to an artificial bimodality in the color magnitude diagram \citep{Sch07, Sal14}.
Therefore, more recent studies tend to exploit a modern definition of ``green-valley'' on the SFR--M$_*$ diagram; i.e. green-valley galaxies are located below the MS by $\sim$0.5--1.5 dex on the SFR--M$_*$ diagram, reflecting the lower SF activity than MS galaxies.
As demonstrated by recent extensive $^{12}$CO($J=1-0$) (hereafter CO) observations, the molecular gas mass fraction ($f_\mathrm{H_2} = M_\mathrm{H_2}/M_*$) and star formation efficiency (${\it SFE} = {\it SFR}/M_\mathrm{H_2}$) both change across the MS \citep{Sai12, Gen15, Sai16}, and accordingly green-valley galaxies are known to have intermediate levels of $f_\mathrm{H_2}$ and SFE on average. A recent study based on the spatially resolved observations of three green-valley galaxies with ALMA and MaNGA by \citet{Lin17} has also confirmed this trend. 

It is known that the majority of green-valley galaxies have larger bulge and smaller disk components compared with normal MS galaxies in general \citep{Bel18, Bre18}, but at the same time, it is also true that there still remains a wide variety of morphologies in green-valley galaxy population; i.e.\ from late-types to early-types \citep{Sch14}. \citet{Sch14} proposed that these green-valley galaxies with early- and late-type morphologies are in different quenching processes.
It is therefore expected that such ``disk+green" and ``bulge+green" galaxies are short-lived (hence rare), but extremely important targets to understand the effect of morphologies during the phase of galaxy transition events. They are equally quenched (judging from their location on the SFR-M$_*$ diagram), but are in the very different stage in terms of their morphological transformation. ``Disk+green" galaxies are in the transitional phase {\it without} starting their morphological change. ``Bulge+green" galaxies are also in the transitional phase but {\it having already completed} their morphological transformation.
It is therefore desirable to directly compare the molecular gas properties in these two distinct population at fixed SFR and M$_*$ to assess the real effect of galaxy morphologies on SF quenching process.

\begin{figure*}[t]
\plottwo{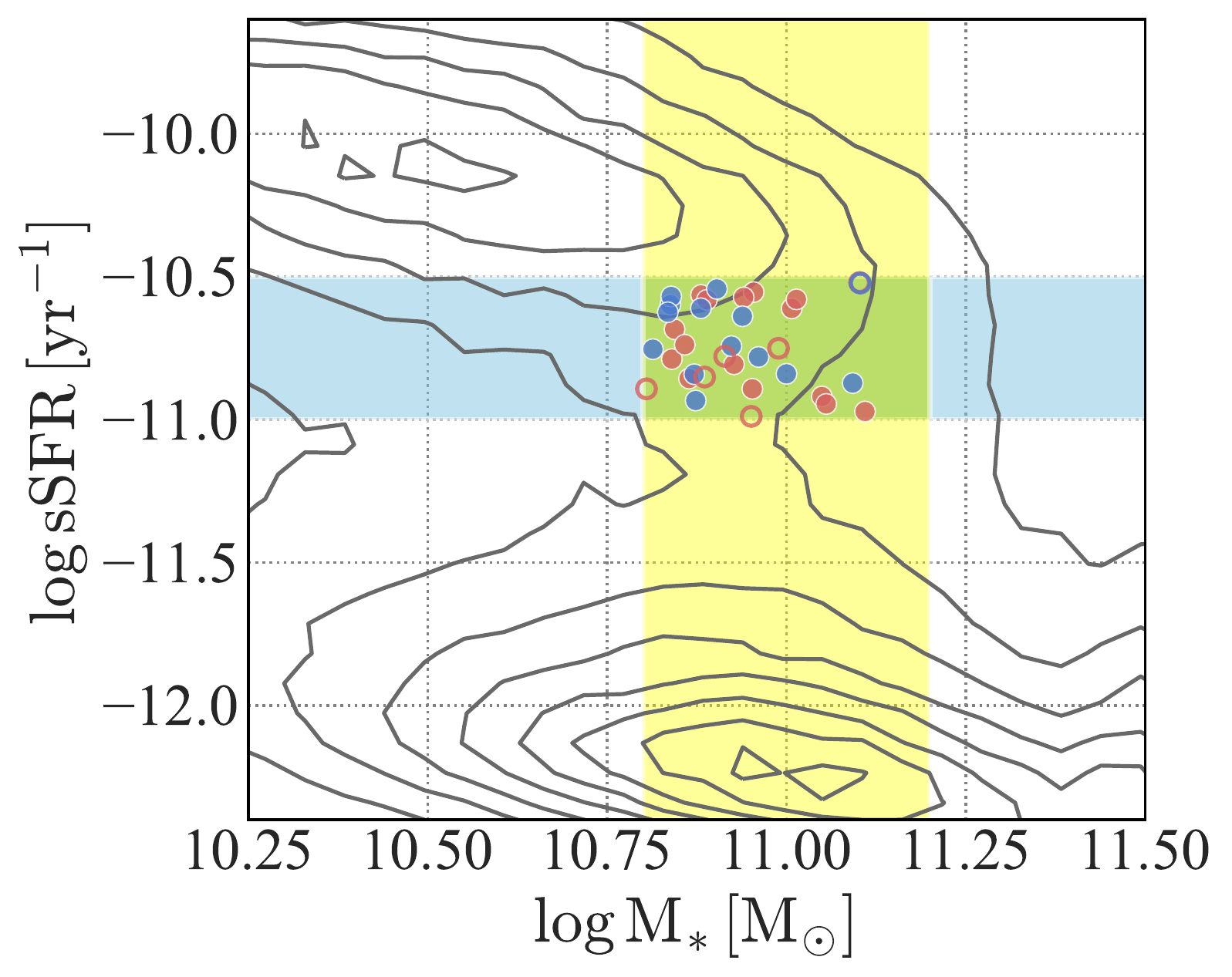}{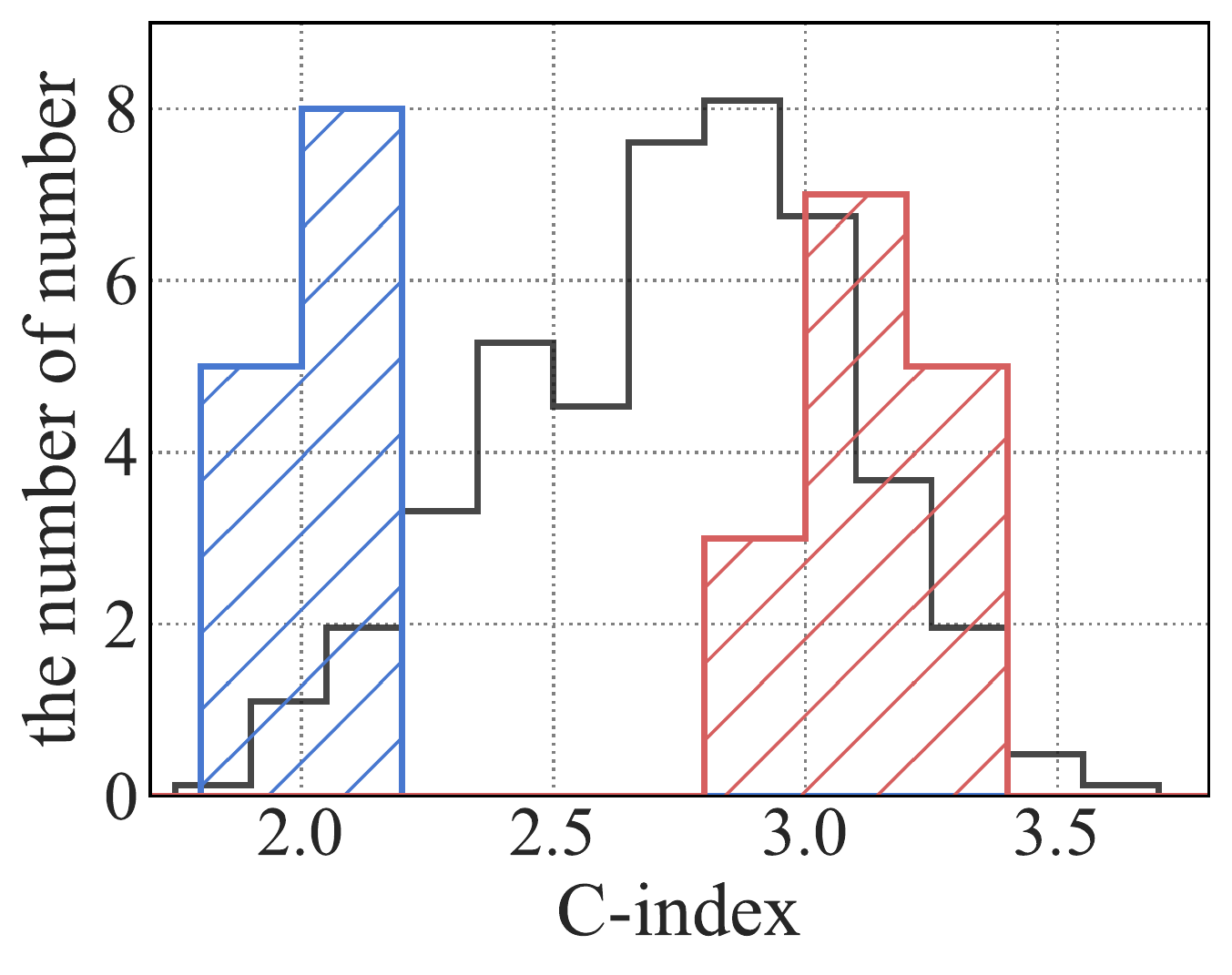}
\caption{
(Left): The stellar mass versus specific SFRs diagram for all SDSS galaxies (gray contours), while the blue and red symbols show the selected disk- and bulge-dominated green-valley galaxies (``disk+green" and ``bulge+green", respectively), where the filled and open circles show galaxies from our NRO~45m observations and the xCOLD~GASS survey, respectively. The yellow and cyan shaded regions show the M$_*$ and sSFR range applied to select our NRO~45m CO observation target galaxies ($10.8 < \log (M_*/\mathrm{M_\odot}) < 11.2$ and $-11 < \log (sSFR/\mathrm{yr^{-1}}) < -10.5$).
(Right): The black histogram shows the arbitrary scaled distribution of C-index for all the SDSS sample with $0.025<z<0.05$, $10.8 < \log (M_*/\mathrm{M_\odot}) < 11.2$ and $-11 < sSFR/\mathrm{yr^{-1}} < -10.5$. The blue and red histograms show the distribution of disk-dominated and bulge-dominated green-valley galaxies targeted by our NRO~45m observations, respectively.
}
\label{sample}
\end{figure*}

In this paper, we present our new CO observations of nearby green-valley galaxies with disk- and bulge-dominated morphologies using the Nobeyama 45m radio telescope (NRO~45m).
By combining our NRO 45m data and other CO data taken from the literature, we directly measure their SFE and test if morphological transformation (i.e.\ the growth of stellar bulge) can really reduce the efficiency of star formation within the galaxies.
This paper is organized as follows.
In Section~\ref{sec:data}, we describe our sample selection and our new CO observations with NRO~45m telescope, as well as our methodology and supplementary data from the xCOLD~GASS survey.
We present our main results on the morphological dependence of the molecular gas properties in Section~\ref{sec:results}, and the discussion in Section~\ref{sec:discussion}.
Finally, our conclusions are presented in Section~\ref{sec:conclusion}.
Throughout this paper, we assume the $\Lambda$CDM universe with $H_0 = 70~\mathrm{km~s^{-1}~Mpc^{-1}}$, $\Omega_m = 0.3$ and $\Omega_{\Lambda} = 0.7$, and \citet{Kro01} initial mass function (IMF).

%%%%%%%%%%  END: INTRODUCTION  %%%%%%%%%%

%%%%%%%%%%  START: DATA  %%%%%%%%%%
\section{data and methodology} \label{sec:data}

\begin{figure*}[htb]
\centering
\includegraphics[width=140mm]{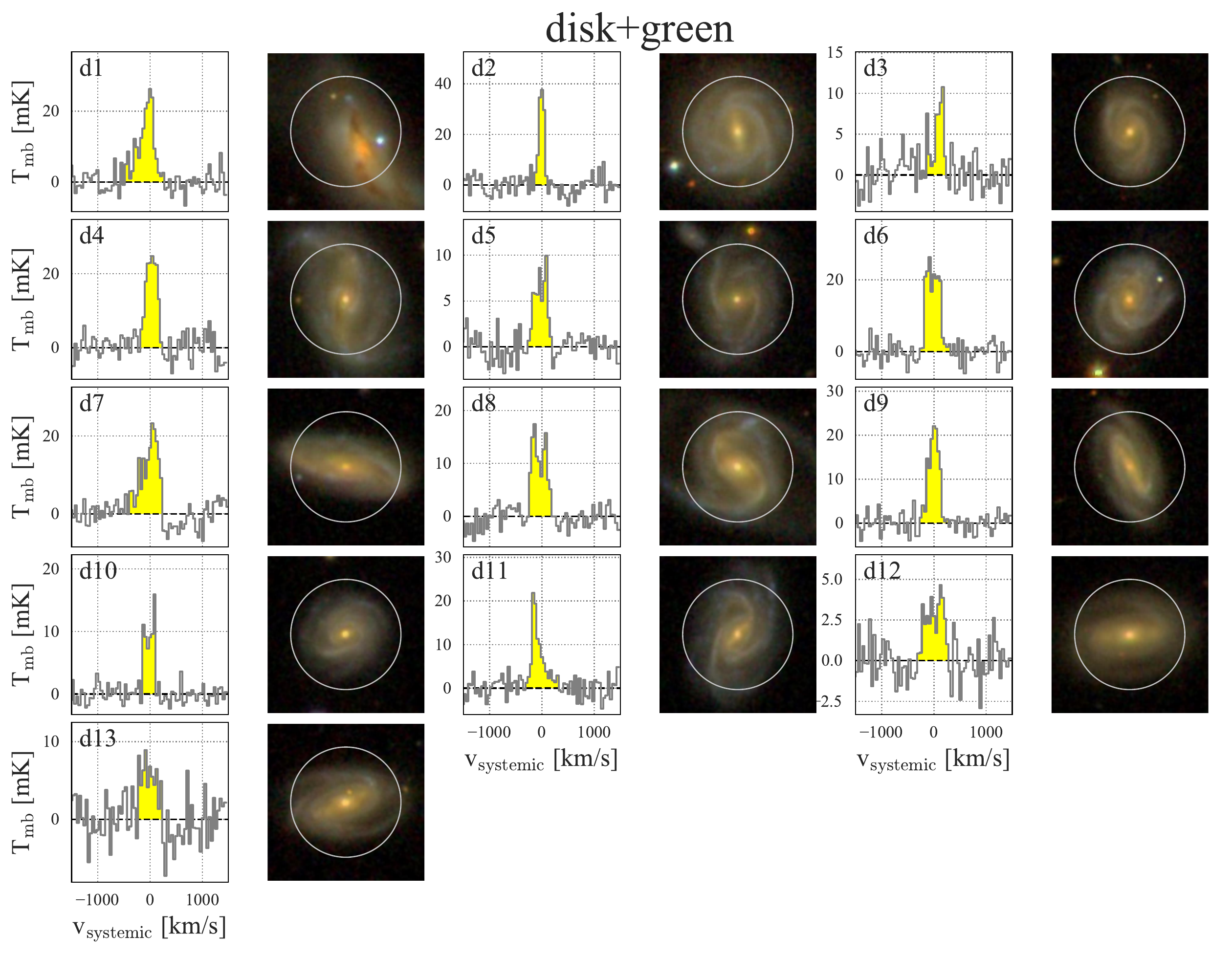}
\includegraphics[width=140mm]{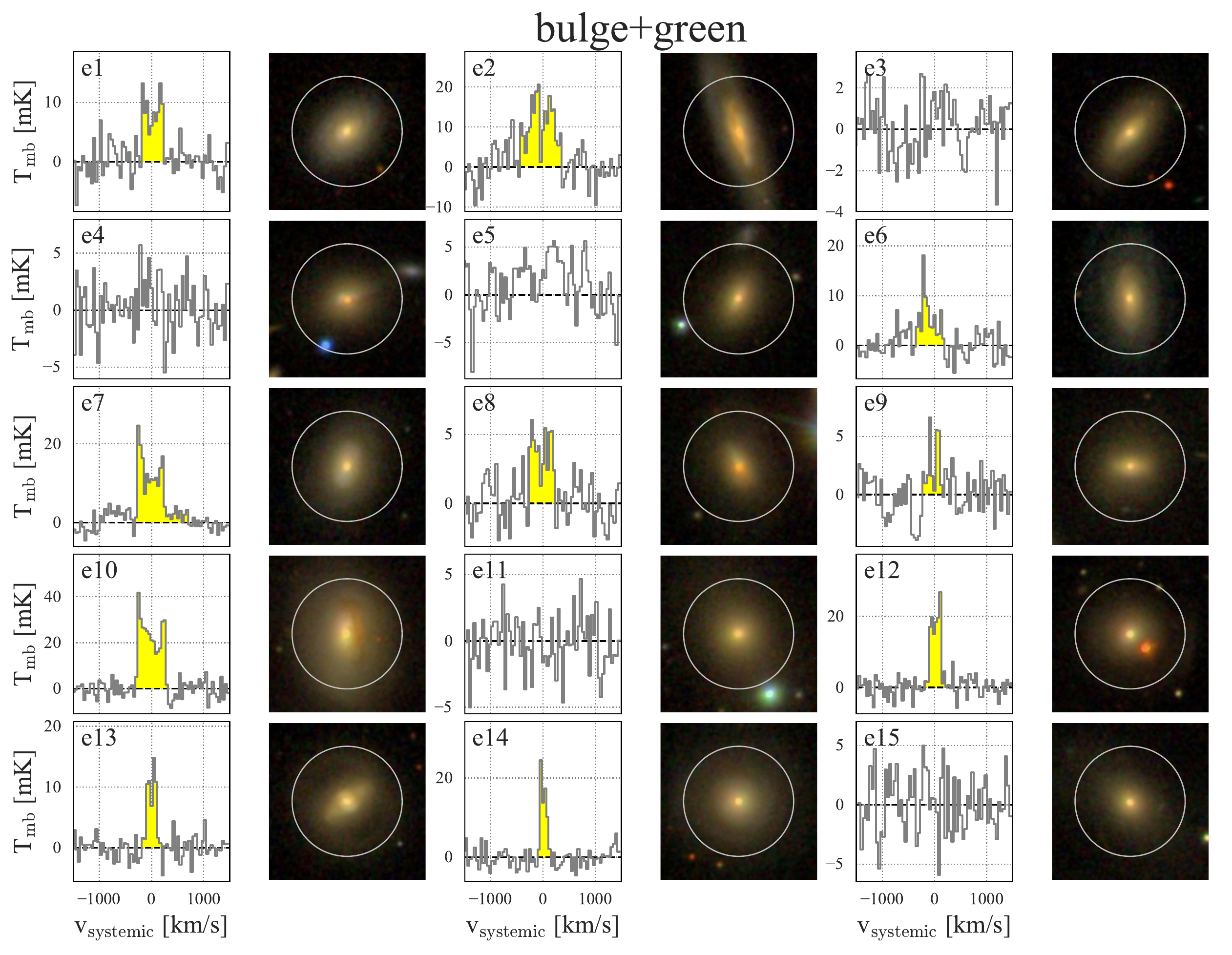}
\caption{The CO spectra obtained by our NRO 45m observations and the SDSS cutout 3-color images (\textit{g, r, i}) for the disk+green (top panels) and bulge+green galaxies (bottom panels). The CO flux of each galaxy is calculated by integrating the flux within the velocity range colored in yellow.
The white circles in the SDSS images show the beam size of NRO 45m telescope ($18 ''$), with its center indicating the SDSS fiber position.
Physical properties of all the target galaxies are summarized in Table \ref{SP}.}
\label{spectra}
\end{figure*}

\subsection{Sample Selection} \label{subsec:sample}
The goal of this paper is to study the morphological dependence of SFE of galaxies in the phase of galaxy transition at fixed M$_*$ and SFR. To achieve this goal, we perform new CO observations of galaxies selected from a small window on the SFR--M$_*$ plane.
We start from the SDSS DR7 spectroscopic catalog \citep{Yor00, Aba09}. M$_*$ and SFR are computed by Max Plank Institute for Astrophysics and Johns Hopkins University (MPA/JHU) group \citep{Kau03, Bri04}, where the SFRs for star-forming galaxies are based either on model fitting to the continuum + emission lines, or on H$_\alpha$ emission line. Therefore, the associated timescale of star formation activity should be short ($\sim$10~Myr), which is relevant for the investigation of star formation activity within the galaxies in the phase of transition.
We then select the target galaxies for our CO observation with NRO~45m telescope following the procedure as described below:

We restrict the redshift range to $0.025 < z < 0.05$. The lower limit ($z>0.025$) is applied to cover the major part of galaxies by a single beam size of NRO~45m telescope ($\sim$18~arcsec at 110~GHz, which corresponds to 9.2~kpc at $z=0.025$). The upper limit ($z<0.05$) is also applied to detect their CO line with a reasonable observational time of the NRO~45m telescope.

To select galaxies in the galaxy transition phase, we select galaxies from a small window on the M--sSFR plane ($10.8 < \log (M_*/\mathrm{M_\odot}) < 11.2$ and $-11 < \log (sSFR/\mathrm{yr^{-1}}) < -10.5$), as shown in the color shades of Figure~\ref{sample}-left.
 
AGNs may enhance the central luminosity of galaxies and affect their morphological classification. It is also expected that the presence of AGNs may affect the star formation activity in the galaxies (AGN feedback). We therefore remove AGNs from our sample based on the SDSS spectral subclass (\texttt{AGN} and/or \texttt{BROADLINE}). 
\footnote{\texttt{AGN} subclass denotes galaxies whose emission line ratios are consistent with the Seyfert or LINER galaxies, and \texttt{BROADLINE} subclass is defined as those having large velocity dispersion ($>$200~km~s$^{-1}$) measured with their emission lines (detected at 10$\sigma$ level).} 
We note that we cross-match our final sample to the {\it swift}/BAT 105-month catalog \citep{Oh18}, and confirm that there is no X-ray source in our sample. 

We compute the concentration index \citep[C-index;][]{Mor58}, defined by $R90_{petro,r}/R50_{petro,r}$, where $R90_{petro,r}$ and $R50_{petro,r}$ are the radius containing 90- and 50-\% of Petrosian flux for SDSS \textit{r}-band photometric data, respectively.
The C-index is known to be strongly correlated with the dominance of bulge component in galaxies \citep{Shi01, Str01}, and we use C-index as the indicator of galaxy morphologies in this study. The black histogram in Figure~\ref{sample}-right shows the (arbitrary scaled) distribution of C-index for all SDSS galaxies with $0.025 < z < 0.05$, $10.8 < \log (M_*/\mathrm{M_\odot}) < 11.2$ and $-11 < \log (sSFR/\mathrm{yr^{-1}}) < -10.5$; i.e.\ the same criteria that we applied to select our NRO~45m target as described above. This plot demonstrates that green-valley galaxies have a wide variety of morphologies even if we fix stellar mass and SFR.   

By visually inspecting the individual galaxies showing high/low C-index, and by considering their target visibility from the NRO~45m telescope, we select 13 disk-dominated green-valley galaxies with C-index$<2.2$ (hereafter ``disk+green'' sample; see the blue histogram in Figure~\ref{sample}-right) and 15 bulge-dominated green-valley galaxies with C-index$>2.8$ (hereafter ``bulge+green'' sample; see the red histogram in Figure~\ref{sample}-right).
We show the SDSS optical color composite image of individual galaxies in Figure~\ref{spectra}, demonstrating that the above procedure allows us to successfully select green-valley galaxies with distinct morphologies.
We note that, in Section~\ref{subsec:Mh2}, we also check the confidence in the C-index as a morphological indicator by comparing it to the stellar mass surface density within 1~kpc.
The blue and red filled circles in Figure~\ref{sample}-left show the distributions of our disk+green and bulge+green galaxies on the sSFR--M$_*$ plane.
In this figure, we also show the supplementary data from the xCOLD~GASS survey with the open circles (see Section~\ref{subsec:xCG} for details).
The properties of our NRO~45m target galaxies are summarized in Table~\ref{SP}.

\subsection{Observation and Data Reduction} \label{subsec:observation}

We performed CO observations of the 13 disk+green and 15 bulge+green galaxies from December to February in the 2017/2018 semester using the NRO~45m telescope (CG171004: Y. Koyama et al.).
The CO emission line (rest-frame 115.27\,GHz) is shifted to 109.78 -- 112.46\,GHz according to the redshift distribution of our sample ($0.025<z<0.05$).
We used a multi-beam 100 GHz band SIS receiver \citep[FOREST;][]{Min16} and a copy of a part of the FX-type correlator for the Atacama Compact Array \citep[SAM45;][]{Kam12}.
We performed single point observations with two beams \citep[ON-ON mode;][]{Nak13}.
A typical on-source integration time is 2 hours for each galaxy.
The $^{13}$CO($J=1-0$) line of IRC+10216 was observed daily as a standard source for flux calibration.
The flux calibration was performed by the chopper wheel method and the scaling factor (\textit{f}) defined as a flux ratio of the standard source and the reference data listed on the NRO website\footnote{\url{https://www.nro.nao.ac.jp/~nro45mrt/html/obs/standard/}}.
The main beam temperature ($T_{\rm mb}$) is calculated by 
\begin{equation}
T_{\rm mb} = T^{*}_a / \eta_{\rm mb},
\end{equation}
where $T^{*}_a$ is the antenna temperature.
The main beam efficiency ($\eta_{\rm mb}$) during the semester is 0.43 at 110\,GHz according to the NRO website\footnote{\url{https://www.nro.nao.ac.jp/~nro45mrt/html/prop/status/Status_latest.html}}.

\begin{figure}
\plotone{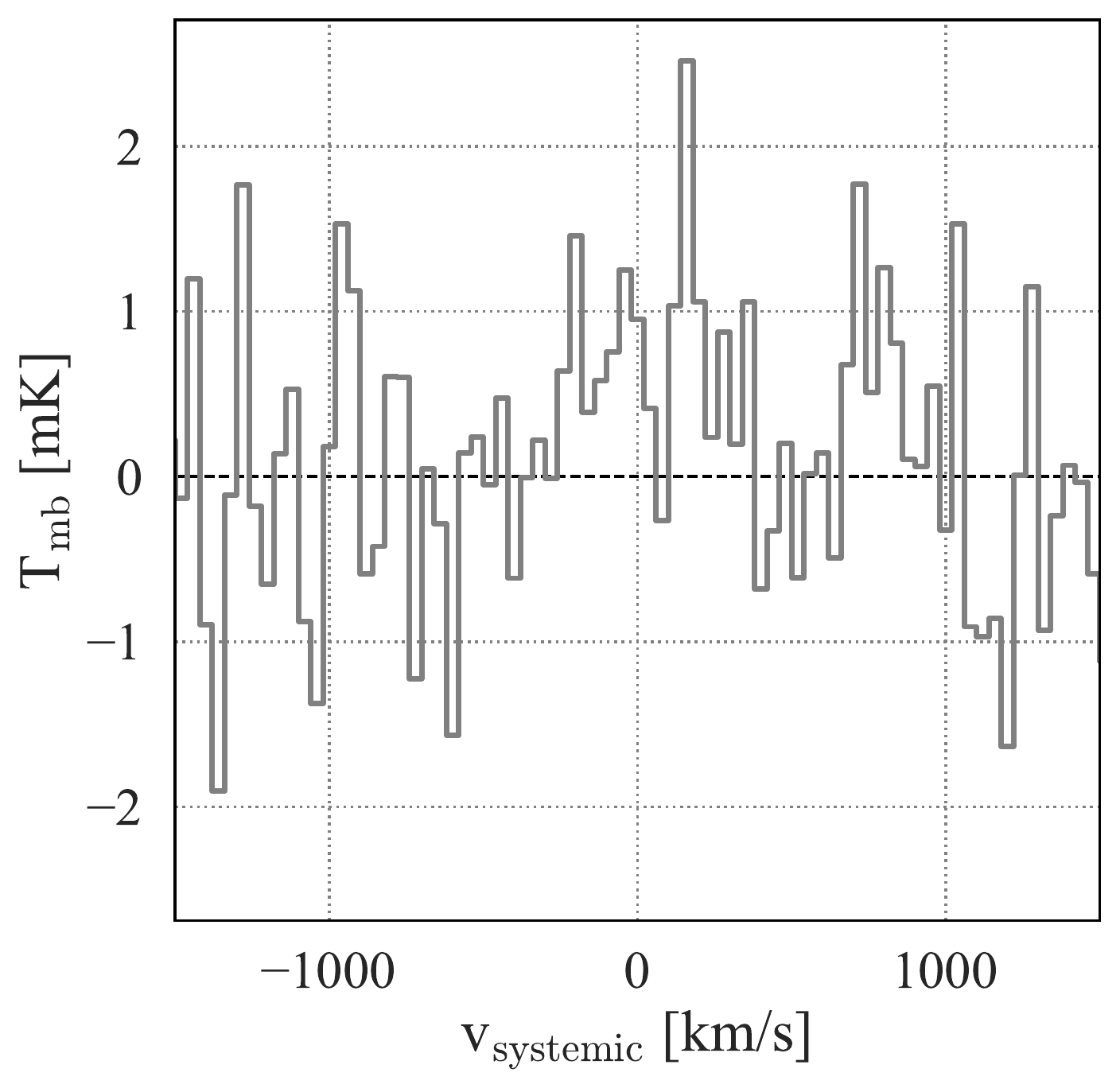}
\caption{The stacked CO spectrum for the five CO-undetected bulge+green galaxies (ID = e3, 4, 5, 11, 15; see Table~\ref{SP}).}
\label{stack}
\end{figure}

\begin{figure*}[t]
\plotone{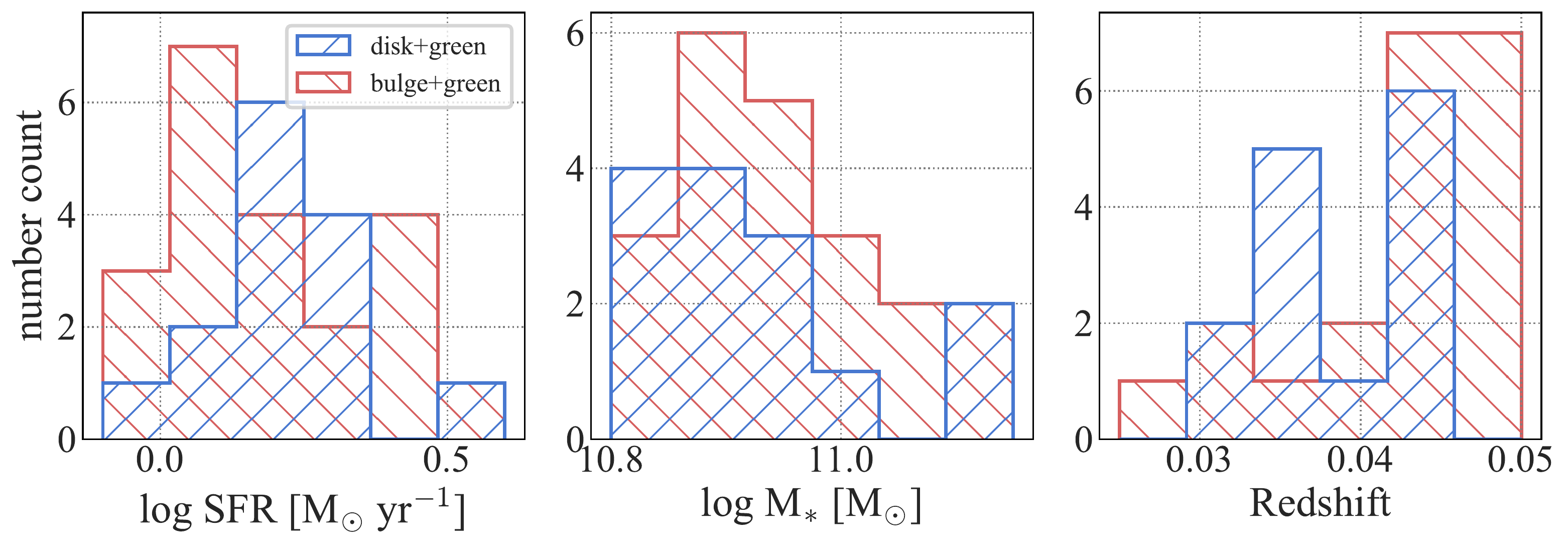}
\caption{The distributions of SFR (left), $M_*$ (middle) and redshift (right) for the disk+green (blue) and bulge+green (red) galaxies. The plots demonstrate that there is no systematic differences between the disk+green and bulge+green galaxies in the distributions of these physical parameters.}
\label{params}
\end{figure*}

We performed data reduction by following the procedure described in \citet{Koy17} using the NEWSTAR software developed by NRO based on the Astronomical Image Processing System (AIPS) package.
We only used the data with wind velocities of $<5\,\rm m\,s^{-1}$, pointing accuracies with $<$$5\arcsec$, system noise temperature ($T_\mathrm{sys}$) with $<$300\,K, and the data with the rms noise temperature of $T_\mathrm{rms}<0.045\,\mathrm{K}$ in the $T_\mathrm{mb}$ scale at a velocity resolution of 200\,km\,s$^{-1}$ to exclude bad baseline spectra.
We subtracted baselines by linear fitting, and combined the spectra for both beams and polarizations. Finally, we performed spectral binning to 40\,km\,s$^{-1}$ resolution.
The observed CO spectra are shown in Figure~\ref{spectra}.
We then calculated the integrated intensity $I_{\rm CO}$ with 
\begin{equation}
I_{\rm CO} = \int T_{\rm mb}\,dv.
\end{equation}
The $T_{\rm rms}$ at the velocity resolution of $\Delta v = 40\,\rm km\,s^{-1}$ is 1.2--4.0\,mK ($T_{\rm mb}$) for our targets.
We detected CO emission line with the signal (peak temperature) -to-noise (rms) ratio (S/N) of $>3$ from 23 galaxies out of 28 observed galaxies.

Finally, we calculated the CO luminosity ($L^{\prime}_{\rm CO}$) of individual galaxies using the following equation:
\begin{equation}
L'_\mathrm{CO} = \frac{{\it\Omega}_b I_\mathrm{CO} D^2_{L}}{(1+z)^3}~\mathrm{[K~km~s^{-1}~pc^2]},
\end{equation}
where $\it\Omega_{\rm b}$ is the beam solid angle of $\theta_{\rm mb}\ (= 18\arcsec)$, and $D_{L}$ is the luminosity distance of each target galaxy.
The physical quantities of our target galaxies derived from our NRO~45m observations are summarized in Table \ref{SP}.
We note that some of the galaxies are slightly larger in optical size than the NRO beam size (white circles in the SDSS images), particularly for the disk+green galaxies.
However, we do not apply aperture correction, because molecular gas component is reported to concentrate on the central part of galaxies \citep[e.g.][]{You91}, and therefore, the effect of CO flux loss due to this aperture effect is expected to be negligibly small.
For five galaxies without CO line detection, we assigned the 3$\sigma$ upper limits of CO luminosities assuming the FWHM of 200\,km s$^{-1}$ (estimated from the mean CO line width of the CO-detected galaxies).

We also performed spectral stacking analysis to estimate the mean CO luminosity of the five CO-undetected bulge+green galaxies (e3, 4, 5, 11, 15). 
The stacking analysis employed the weight mean stacking for the spectra with:
\begin{equation}
T_{\mathrm{mb}, stacked} = \frac{\Sigma_{i} w_{i}T_{mb, i}}{\Sigma_{i}w_{i}},
\end{equation}
where \textit{w} denotes $1/rms^{2}$ for each spectrum. We show the stacked spectrum in Figure~\ref{stack}.
Although $T_\mathrm{rms}$ is properly reduced by the stacking analysis ($= 0.89\,\mathrm{mK}$), CO line is still undetectable at a S/N$>3$ level.
We therefore assign their (average) 3$\sigma$ upper limit from the noise level of this stacking analysis. We also report the mean properties of the CO-undetected bulge+green galaxies from this stacking analysis at the bottom line of Table \ref{SP}.

\movetabledown=2.1in
\begin{table*}[p]
\begin{rotatetable*}
\caption{Summary of the physical quantities of our NRO~45m green-valley galaxy sample. \label{SP}}
{\scriptsize
\begin{tabular}{rrrrrrrrrrrr} \hline \hline
ID & Morphology & C-index & RA & Dec & {\it z} & log SFR$_\mathrm{MPA/JHU}$ & log SFR$_{\it WISE}$ & log M$_*$ & I$_\mathrm{CO}$ & log L$_\mathrm{CO}$ & log M$_\mathrm{H_2}$ \\
 & & & (hms) & (dms) & & (M$_\odot$ yr$^{-1}$) & (M$_\odot$ yr$^{-1}$) & (M$_\odot$) & (K km s$^{-1}$) & (K km s$^{-1}$pc$^2$) & (M$_\odot$) \\ \hline
d1   & disk & 1.80 & 08 04 46.760 & +10 46 41.83 & 0.035 & -0.06$_{-0.34}^{+0.39}$ & 0.81$\pm$0.14 & 10.87$_{-0.09}^{+0.10}$ & 7.78$\pm$2.13 & 9.17$\pm$0.12 & 9.81$\pm$0.12 \\
d2   & disk & 1.90 & 13 50 32.134 & +03 11 38.96 & 0.032 & 0.24$_{-0.45}^{+0.39}$ & 0.43$\pm$0.14 & 10.84$_{-0.10}^{+0.09}$ & 4.72$\pm$0.87 & 8.88$\pm$0.08 & 9.52$\pm$0.08 \\
d3   & disk & 1.92 & 17 11 47.781 & +28 19 27.27 & 0.042 & 0.06$_{-0.56}^{+0.42}$ & 0.24$\pm$0.14 & 10.81$_{-0.09}^{+0.10}$ & 1.75$\pm$0.74 & 8.68$\pm$0.18 & 9.31$\pm$0.18 \\
d4   & disk & 1.95 & 13 13 13.457 & +33 59 04.00 & 0.035 & 0.18$_{-0.39}^{+0.38}$ & 0.60$\pm$0.14 & 10.96$_{-0.10}^{+0.10}$ & 6.92$\pm$1.18 & 9.12$\pm$0.07 & 9.75$\pm$0.07 \\
d5   & disk & 1.95 & 10 25 17.750 & +17 08 21.09 & 0.045 & 0.03$_{-0.34}^{+0.36}$ & 0.18$\pm$0.14 & 10.87$_{-0.10}^{+0.09}$ & 2.26$\pm$0.60 & 8.83$\pm$0.12 & 9.47$\pm$0.12 \\
d6   & disk & 2.04 & 13 22 40.387 & +32 53 22.76 & 0.036 & 0.27$_{-1.32}^{+0.80}$ & 0.58$\pm$0.14 & 10.88$_{-0.12}^{+0.13}$ & 7.47$\pm$1.53 & 9.17$\pm$0.09 & 9.80$\pm$0.09 \\
d7   & disk & 2.07 & 15 06 34.472 & +05 13 25.12 & 0.035 & 0.27$_{-0.32}^{+0.35}$ & 0.50$\pm$0.14 & 10.84$_{-0.09}^{+0.10}$ & 7.30$\pm$1.31 & 9.13$\pm$0.08 & 9.76$\pm$0.08 \\
d8   & disk & 2.09 & 11 45 17.564 & +26 46 02.63 & 0.030 & 0.30$_{-0.35}^{+0.35}$ & 0.45$\pm$0.14 & 10.94$_{-0.09}^{+0.09}$ & 4.51$\pm$0.64 & 8.81$\pm$0.06 & 9.44$\pm$0.06 \\
d9   & disk & 2.09 & 12 26 32.710 & +16 50 40.66 & 0.045 & 0.36$_{-1.08}^{+0.53}$ & 0.75$\pm$0.14 & 10.90$_{-0.09}^{+0.10}$ & 5.20$\pm$0.95 & 9.20$\pm$0.08 & 9.84$\pm$0.08 \\
d10 & disk & 2.11 & 08 43 06.123 & +25 32 20.62 & 0.044 & 0.21$_{-0.57}^{+0.44}$ & 0.24$\pm$0.14 & 10.83$_{-0.09}^{+0.10}$ & 2.52$\pm$0.33 & 8.87$\pm$0.06 & 9.51$\pm$0.06 \\
d11 & disk & 2.12 & 09 48 54.495 & +24 52 29.04 & 0.043 & 0.16$_{-0.29}^{+0.36}$ & 0.56$\pm$0.14 & 11.00$_{-0.11}^{+0.12}$ & 4.08$\pm$1.36 & 9.07$\pm$0.15 & 9.70$\pm$0.15 \\
d12 & disk & 2.15 & 12 52 44.495 & +59 15 56.47 & 0.043 & 0.22$_{-0.33}^{+0.31}$ & 0.09$\pm$0.14 & 11.09$_{-0.09}^{+0.09}$ & 1.48$\pm$0.68 & 8.62$\pm$0.20 & 9.25$\pm$0.20 \\
d13 & disk & 2.16 & 15 37 03.843 & +25 33 55.25 & 0.035 & 0.18$_{-0.31}^{+0.32}$ & 0.24$\pm$0.14 & 10.92$_{-0.09}^{+0.10}$ & 2.90$\pm$1.10 & 8.73$\pm$0.16 & 9.36$\pm$0.16 \\
e1   & bulge & 2.84 & 10 44 54.128 & +27 28 08.74 & 0.044 & 0.40$_{-0.43}^{+0.39}$ & 0.36$\pm$0.14 & 10.95$_{-0.09}^{+0.10}$ & 3.64$\pm$1.18 & 9.04$\pm$0.14 & 9.67$\pm$0.14 \\
e2   & bulge & 2.96 & 13 01 07.398 & +54 47 57.42 & 0.031 & 0.40$_{-1.08}^{+0.51}$ & 0.67$\pm$0.14 & 11.01$_{-0.10}^{+0.10}$ & 9.55$\pm$2.94 & 9.16$\pm$0.13 & 9.80$\pm$0.13 \\
e3   & bulge & 2.99 & 12 30 55.791 & +51 16 56.59 & 0.043 & 0.01$_{-1.28}^{+0.59}$ & -0.11$\pm$0.14 & 10.86$_{-0.09}^{+0.09}$ & $<$ 1.93 & $<$ 8.74 & $<$ 9.37 \\
e4   & bulge & 3.02 & 14 38 56.819 & +28 21 11.76 & 0.044 & 0.32$_{-0.27}^{+0.34}$ & 0.00$\pm$0.14 & 10.88$_{-0.09}^{+0.10}$ & $<$ 2.89 & $<$ 8.94 & $<$ 9.57 \\
e5   & bulge & 3.02 & 08 38 45.730 & +25 14 11.27 & 0.044 & 0.05$_{-0.48}^{+0.41}$ & $<$ -0.51 & 10.84$_{-0.09}^{+0.10}$ & $<$ 3.53 & $<$ 9.01 & $<$ 9.64 \\
e6   & bulge & 3.04 & 08 16 44.864 & +27 35 30.76 & 0.040 & 0.31$_{-0.39}^{+0.38}$ & 0.21$\pm$0.14 & 10.89$_{-0.09}^{+0.10}$ & 3.21$\pm$1.27 & 8.89$\pm$0.17 & 9.53$\pm$0.17 \\
e7   & bulge & 3.07 & 13 34 09.413 & +13 16 50.94 & 0.044 & 0.43$_{-0.33}^{+0.34}$ & 0.53$\pm$0.14 & 11.01$_{-0.09}^{+0.10}$ & 6.85$\pm$1.27 & 9.31$\pm$0.08 & 9.94$\pm$0.08 \\
e8   & bulge & 3.07 & 17 02 37.467 & +24 52 10.02 & 0.049 & 0.14$_{-1.24}^{+0.65}$ & 0.19$\pm$0.14 & 11.11$_{-0.09}^{+0.10}$ & 1.68$\pm$0.76 & 8.78$\pm$0.20 & 9.42$\pm$0.20 \\
e9   & bulge & 3.10 & 11 16 32.347 & +29 16 33.46 & 0.046 & 0.37$_{-0.25}^{+0.32}$ & $<$ -0.22 & 10.94$_{-0.09}^{+0.10}$ & 1.25$\pm$0.76 & 8.61$\pm$0.26 & 9.24$\pm$0.26 \\
e10 & bulge & 3.13 & 10 22 19.386 & +36 34 58.90 & 0.026 & 0.06$_{-0.63}^{+0.41}$ & 0.47$\pm$0.14 & 10.95$_{-0.09}^{+0.10}$ & 13.04$\pm$2.05 & 9.13$\pm$0.07 & 9.77$\pm$0.07 \\
e11 & bulge & 3.28 & 11 16 23.197 & +12 00 55.32 & 0.046 & 0.13$_{-1.15}^{+0.55}$ & $<$ -0.16 & 11.05$_{-0.09}^{+0.09}$ & $<$ 2.89 & $<$ 8.98 & $<$ 9.61 \\
e12 & bulge & 3.31 & 16 26 30.965 & +25 53 40.53 & 0.050 & 0.16$_{-0.59}^{+0.38}$ & 0.50$\pm$0.14 & 10.84$_{-0.09}^{+0.10}$ & 5.03$\pm$0.91 & 9.28$\pm$0.08 & 9.91$\pm$0.08 \\
e13 & bulge & 3.32 & 14 52 32.147 & +17 03 46.03 & 0.045 & 0.11$_{-0.95}^{+0.47}$ & 0.11$\pm$0.14 & 11.06$_{-0.10}^{+0.09}$ & 2.33$\pm$0.57 & 8.85$\pm$0.11 & 9.48$\pm$0.11 \\
e14 & bulge & 3.33 & 15 37 26.092 & +21 44 37.68 & 0.041 & 0.12$_{-0.71}^{+0.40}$ & 0.31$\pm$0.14 & 10.93$_{-0.10}^{+0.09}$ & 2.78$\pm$0.55 & 8.84$\pm$0.09 & 9.48$\pm$0.09 \\
e15 & bulge & 3.34 & 14 33 12.953 & +52 57 47.56 & 0.047 & 0.12$_{-0.36}^{+0.36}$ & -0.09$\pm$0.14 & 10.86$_{-0.09}^{+0.10}$ & $<$ 3.21 & $<$ 9.04 & $<$ 9.67 \\ \hline
$\langle e3,4,5,11,15 \rangle$ & bulge & 3.13 & --- & --- & 0.045 & 0.13 & -0.09 & 10.90 & $<$ 1.26 & $<$ 8.59 & $<$ 9.22 \\ \hline
\multicolumn{12}{l}{Note--log SFR$_{\it WISE}$, I$_\mathrm{CO}$, L$_\mathrm{CO}$ and M$_\mathrm{H_2}$ with "$<$" denote the upper limits.}
\end{tabular}}
\end{rotatetable*}
\end{table*}

\subsection{xCOLD GASS} \label{subsec:xCG}

In this study, we also use the CO data of local galaxies publicly available from the xCOLD GASS survey \citep{Sai17}.
The xCOLD~GASS is an extended version of the COLD~GASS survey \citep{Sai11a, Sai11b, Sai12} performed with the IRAM~30m telescope, and provides the most extensive CO datasets of 532 local galaxies in a redshift range of $0.01 < z < 0.05$.
 
We measured their C-index applying the same criteria that we used to select our NRO~45m targets (see Section~2.1), and then we selected 1 disk+green and 6 bulge+green galaxies from the xCOLD~GASS sample (see the open circles in Figure~\ref{sample}-left). The small number of the xCOLD~GASS sample satisfying our criteria is not surprising because the xCOLD~GASS survey is originally designed to cover a wide range in the M$_*$--SFR plane, whilst our focus is to study galaxies at fixed M$_*$ and SFR.

\subsection{The Final Sample} \label{subsec:final}

By combining our NRO~45m sample and the xCOLD~GASS sample, our final sample includes 14 disk+green and 21 bulge+green galaxies (35 galaxies in total).
We calculate the molecular gas mass (M$_\mathrm{H_2}$) of individual galaxies with $\mathrm{M_{H_2}} = \alpha_\mathrm{CO}\mathrm{L'_{CO}}$, where $\alpha_\mathrm{CO}$ is the CO-to-H$_2$ conversion factor.
We adopt the Galactic value of $\alpha_\mathrm{CO} = 4.3$ M$_\odot\,(\mathrm{K\,km\,s^{-1}\,pc^2})^{-1}$,
which includes the contribution of heavy elements (mainly from helium), as commonly used in studies of star-forming galaxies in the local universe \citep{Bol13}.
We note that the $\alpha_\mathrm{CO}$ value is reported to depend on the gas-phase metallicity of galaxies \citep[e.g.][]{Gen12, Bol13}.
Unfortunately, it is not possible to measure the metallicities for many of our green-valley galaxies due to their weak emission lines, 
but by comparing the average metallicities for 5 of 14 disk+green galaxies ($9.13\pm0.02$) and for 3 of 21 bulge+green galaxies ($9.01\pm0.14$) listed in the MPA-JHU catalog \citep{Tre04}, we conclude that the difference in $\alpha_\mathrm{CO}$ due to the metallicity effect between the disk+green and bulge+green samples must be negligible. 

Figure~\ref{params} shows the M$_*$, SFR, and redshift distributions of our final bulge+green and disk+green galaxies.
As intended, there is no systematic difference between the disk+green and bulge+green galaxies in the distribution of M$_*$, SFR, and redshift.
We emphasize again that the aim of this paper is to assess the real effect of galaxy morphologies on SF quenching (SF efficiency) at fixed M$_*$ and SFR by eliminating any bias from the analyses.

%%%%%%%%%%  START: RESULTS  %%%%%%%%%%
\section{Results} \label{sec:results}
\subsection{The Distribution of M$_\mathrm{H_2}$, $f_\mathrm{H_2}$ and SFE} \label{subsec:Mh2}

\begin{figure*}
\plotone{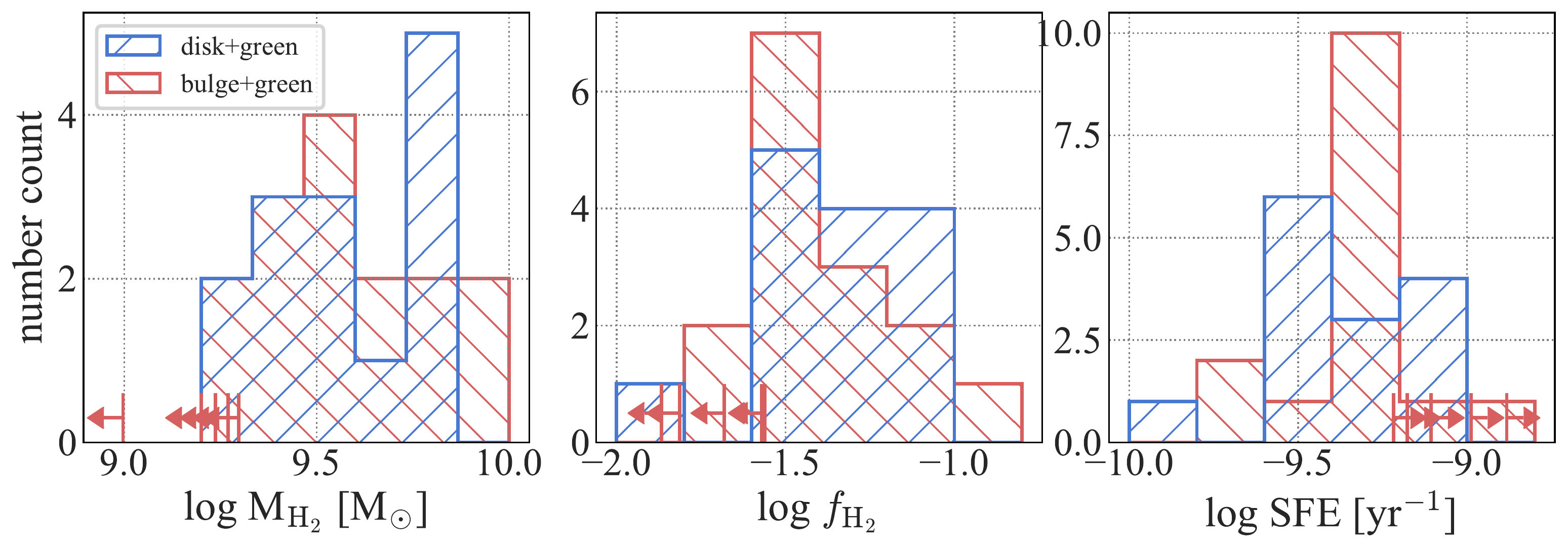}
\caption{The distributions of $M_\mathrm{H_2}$ (left), $f_\mathrm{H_2}$ (middle) and SFE (right) for the disk+green (blue) and bulge+green (red) galaxies.
The color coding is the same as that in Figure~\ref{params}.
The red arrows show the upper/lower limits for the CO-undetected bulge+green galaxies, which are not included in the histograms.}
\label{Mh2}
\end{figure*}

\begin{figure}
\plotone{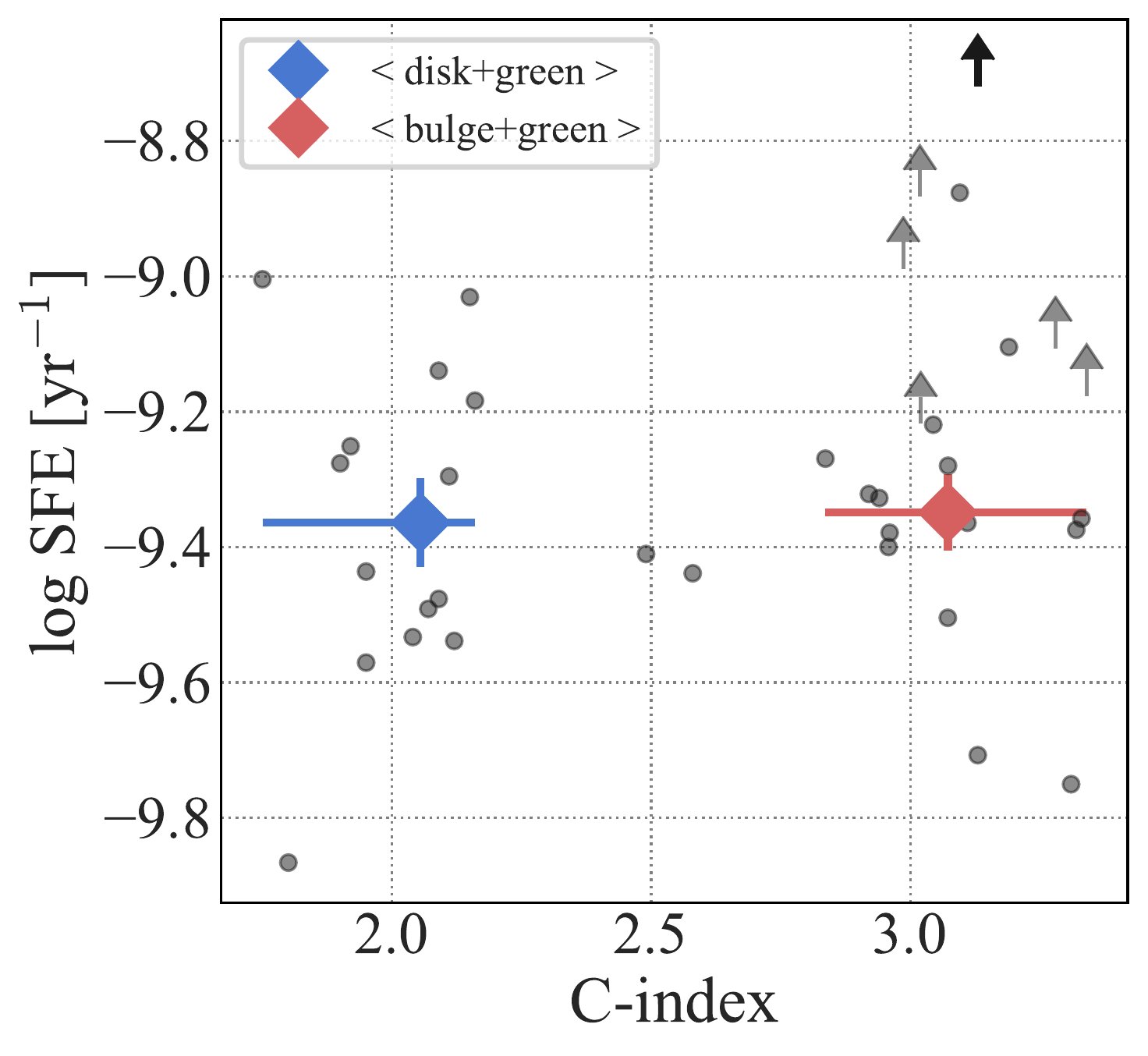}
\caption{The relation between C-index and SFE at fixed $M_*$ and SFR.
For comparison, we also plot two xCOLD GASS sources with intermediate morphology ($2.2 <$ C-index $< 2.8$) in the same $M_*$--sSFR window ($10.8 < \log (M_*/\mathrm{M_\odot}) < 11.2$ and $-11 < \log (sSFR/\mathrm{yr^{-1}}) < -10.5$).
The gray arrows show the upper limits for the individual CO-undetected sources, while the black arrow indicate the upper limit from their stacking analysis.
The blue and red diamonds show the mean SFE for the CO-detected disk+green and bulge+green galaxiess, respectively,
where the error bars for the vertical and horizontal axis represent the standard error of the mean SFE and the range of C-index for each subsample, respectively.
}
\label{C_SFE}
\end{figure}

\begin{deluxetable}{rccc}[t!]
\tablecaption{Means and standard deviations of $M_\mathrm{H_2}$, $f_\mathrm{H_2}$ and SFE for the CO-detected disk+green and bulge+green galaxies. \label{stat}}
\tablewidth{0pt}
\tabletypesize{\scriptsize}
\tablehead{
\colhead{Morphology} &
\colhead{log M$_\mathrm{H_2}$ [M$_\odot$]} &
\colhead{log $f_\mathrm{H_2}$} &
\colhead{log SFE [yr$^{-1}$]}
}
\startdata
disk-dominated & $9.59 \pm  0.20$ & $-1.34 \pm  0.23$ & $-9.36 \pm  0.24$ \\
bulge-dominated &  $9.54 \pm  0.25$ & $-1.41 \pm  0.25$ & $-9.35 \pm  0.22$ \\
\enddata
\end{deluxetable}

We compare the molecular gas properties of green-valley galaxies with different morphological types as a first step.
In Figure~\ref{Mh2}, we show the distribution of $M_\mathrm{H_2}$, $f_\mathrm{H_2}$ and SFE ($= SFR/M_\mathrm{H_2}$) for our disk+green and bulge+green galaxies.
The mean properties of CO-detected galaxies with each morphological type are summarized in Table \ref{stat}.
We perform the two-sample Kolmogorov-Smirnov (KS) test between disk+green and bulge+green galaxies.
The derived {\it p}-values (i.e. the probability that the two samples are drawn from the same parent population) are 0.98 for $M_\mathrm{H_2}$, 0.72 for $f_\mathrm{H_2}$ and 0.45 for SFE, suggesting that there is no clear morphological dependence of $M_\mathrm{H_2}$, $f_\mathrm{H_2}$ and SFE at least for the CO-detected galaxies.

As we reported in Section~\ref{subsec:observation}, there are five CO-undetected galaxies in our bulge+green galaxies. Because we did not detect their CO lines even in the stacked spectrum (see Section~\ref{subsec:observation} and Figure~\ref{stack}), the CO-undetected galaxies might be deficient in molecular gas (in spite of their similar levels of SFRs). 
This result suggests that these galaxies have significantly higher SFE than the other galaxies.

In Figure~\ref{C_SFE}, we plot the SFE against their C-index to more quantitatively examine the morphological dependence of the SFE. It can be seen that there is no significant difference between the CO-detected disk+green (blue diamond) and bulge+green galaxies (red diamond), consistent with the results shown in Figure~\ref{Mh2}, whilst the result from the stacking analysis for the CO-undetected galaxies (shown with the black arrow) seems to show a significantly higher SFE.
This is an opposite trend to that predicted by the morphological quenching scenario, and this population might represent an interesting population to understand the morphological impacts on SF quenching. We will discuss this issue more in detail in Section \ref{subsec:comparison}.

\begin{figure}[h]
\plotone{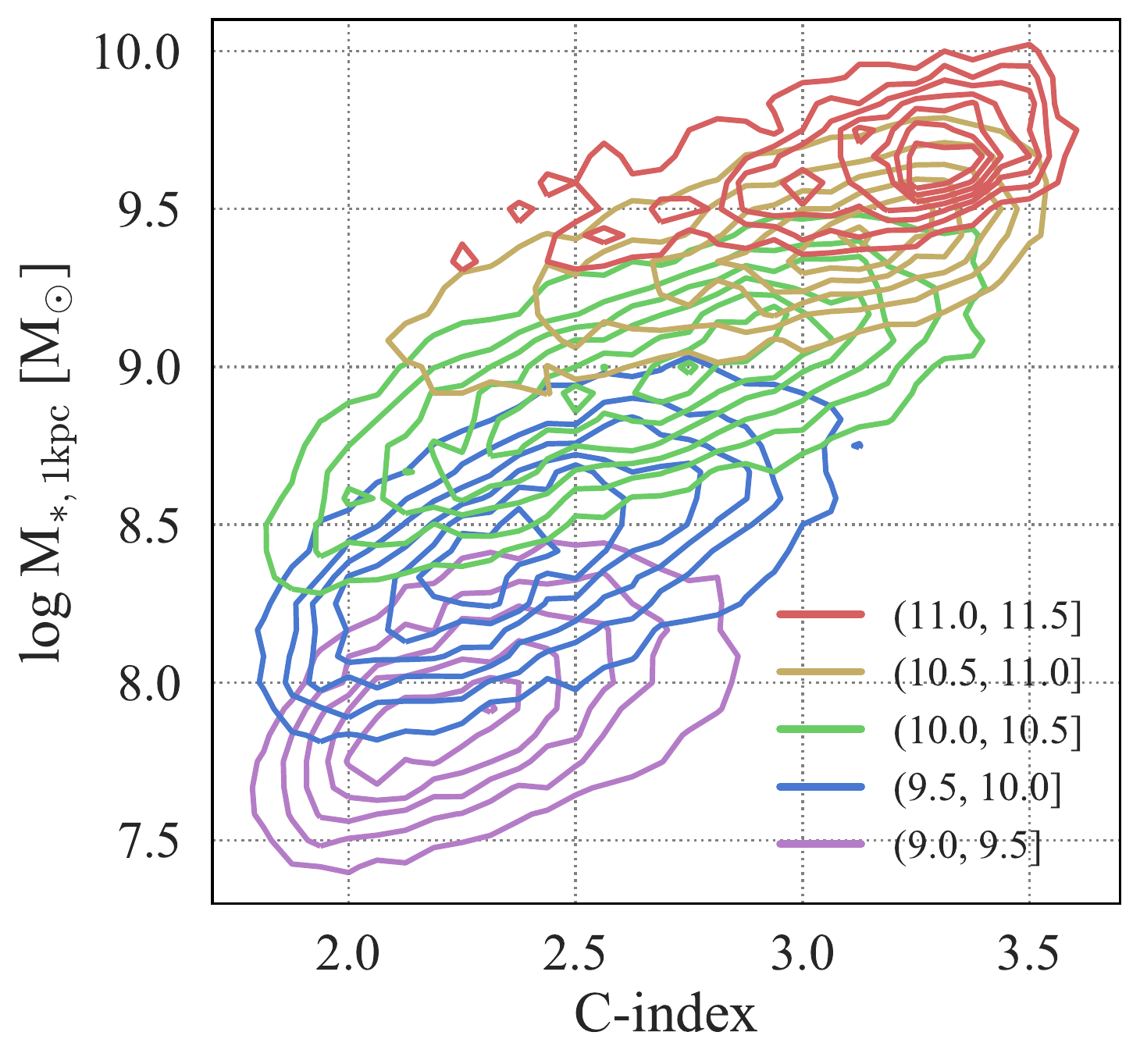}
\caption{
The relation between C-index and M$_\mathrm{*, 1kpc}$ for all the SDSS galaxies with $0.025 < z < 0.05$. The different color contours mean the different $M_*$ range; i.e. $M_*$ increases from purple to red contours as shown in the figure.
There is a tight positive correlation between C-index and M$_\mathrm{*, 1kpc}$ at fixed stellar mass. Since our sample is selected from a small stellar mass range ($10.8 < \log (M_*/\mathrm{M_\odot}) < 11.2$), which is corresponding to yellow and red contours, the C-index is well correlated with the M$_\mathrm{*, 1kpc}$.}
\label{Sigma}
\end{figure}

\begin{figure}[h]
\plotone{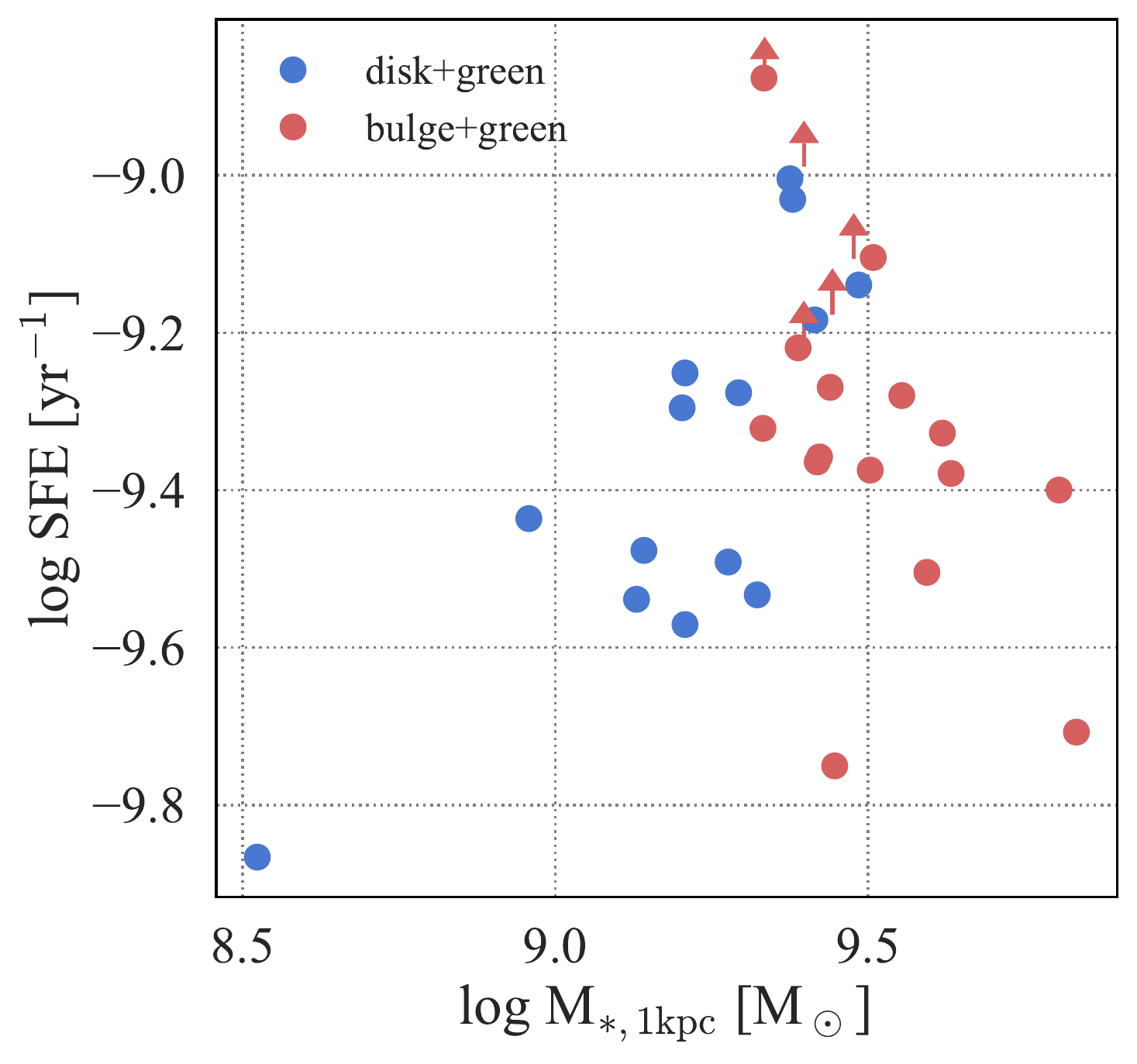}
\caption{
The SFE as a function of $M_\mathrm{*, 1kpc}$.
The disk+green and bulge+green galaxies, defined by C-index in this work, are shown by blue and red symbols.}
\label{Sigma_SFE}
\end{figure}

We note that the C-index is sensitive to the difference of the bulge-to-total stellar mass ratio of galaxies rather than their bulge mass, while some studies suggest that the passive galaxy fraction strongly depends on the bulge mass of galaxies rather than their bulge-to-total stellar mass ratio \citep{Blu14}.
We therefore test the effects of a different morphological indicator, by computing the stellar mass surface density measured within the central 1 kpc ($M_\mathrm{*, 1kpc}$) as another morphological indicator, which better represents the bulge mass of individual galaxies.
We estimate $M_\mathrm{*, 1kpc}$ by using the stellar mass measured in the SDSS $3\arcsec$ fiber (corresponding to $\sim 1.5 - 3.0~\mathrm{kpc}$ at $z=0.025-0.05$) listed in MPA/JHU catalog and by simply scaling it to 1~kpc area.
In Figure~\ref{Sigma}, we show the relation between C-index and $M_\mathrm{*, 1kpc}$ with the color code indicating the difference of stellar mass.
It can be seen that there is a tight positive correlation between C-index and M$_\mathrm{*, 1kpc}$ at fixed stellar mass.
Since our sample is selected from a small stellar mass range ($10.8 < \log (M_*/\mathrm{M_\odot}) < 11.2$), this plot demonstrates that the C-index is well correlated with the M$_\mathrm{*, 1kpc}$ as long as we consider the sample used in this study. In Figure~\ref{Sigma_SFE}, we plot SFE as a function of M$_\mathrm{*, 1kpc}$, with the red and blue symbols indicating the bulge+green and disk+green galaxies classified by C-index, confirming our claim that there is no significant correlation between the SFE and galaxy morphologies. This plot also demonstrates that the disk+green and bulge+green galaxies are also well separated by $M_\mathrm{*, 1kpc}$. We therefore conclude that our results are not affected by the use of different morphological indicators. 
%%%%%%%%%%%%%%%%%%%%%%%%%%%%%%
%%%%%%%%%%%%%%%%%%%%%%%%%%%%%%


\subsection{Morphological Dependence of $M_\mathrm{H_2}$--SFR Relation} \label{subsec:Mh2_SFR}

It is well known that there is a positive correlation between $M_\mathrm{H_2}$ and SFR---this is a fundamental correlation describing star formation \citep[e.g.][]{Ken98}.
We investigate the distribution of disk+green and bulge+green galaxies on the $M_\mathrm{H_2}$--SFR plane.

\begin{figure*}[t]
\plotone{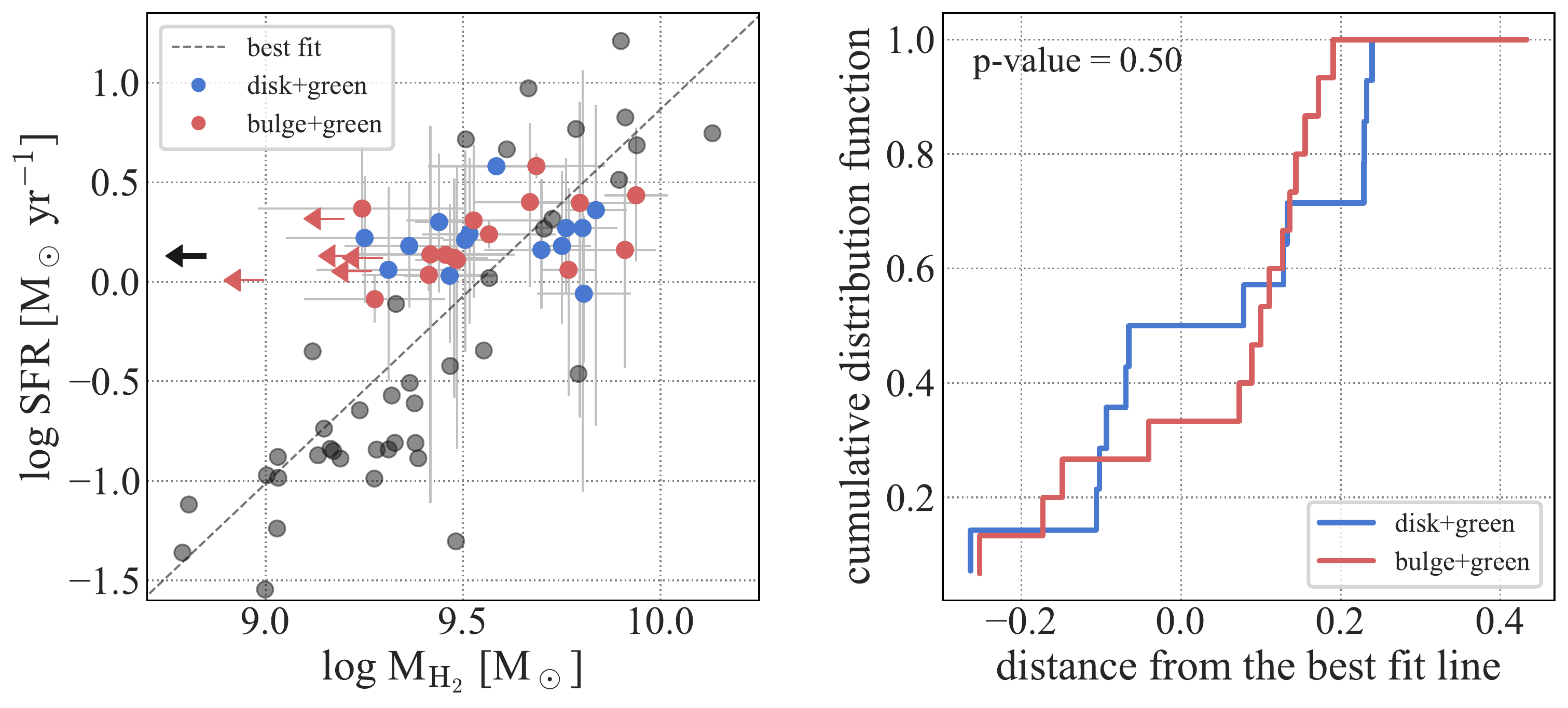}
\caption{(Left): The relation between M$_\mathrm{H_2}$ and SFR for disk+green (blue circles) and bulge-green (red circles) galaxies.
The red arrows show the upper limits of M$_\mathrm{H_2}$ for each CO-undetected bulge+green galaxies, and the black arrow shows the upper limit from their stacking analysis.
The dashed line indicates the best-fit relation determined by using the high-mass sample ($10.8 < \log \mathrm{M_*/M_\odot} < 11.2$) of xCOLD GASS galaxies (gray circles).
(Right): CDFs of the distance in the orthogonal direction to the best-fit line for disk- and bulge-dominated galaxies.
The {\it p}-value derived from two-sample KS test is 0.50, suggesting there is no significant difference between the distribution of the two samples in this diagram. We note that the CO-undetected galaxies are not included in this plot, but it is unlikely that those CO-undetected sources affect our conclusion (see Section~\ref{subsec:Mh2_SFR} for details).}
\label{Mh2_SFR}
\end{figure*}

Figure~\ref{Mh2_SFR}-left shows the M$_\mathrm{H_2}$--SFR relation for our green-valley galaxy samples with different morphologies. A visual inspection suggests that the CO-detected disk+green (blue circles) and bulge+green (red circles) galaxies follow the same relation, or at least the two samples are distributed in the same regions on this diagram. 
To statistically compare the distribution between the disk+green and bulge+green galaxies on the M$_\mathrm{H_2}$--SFR plane, we first fit all the xCOLD~GASS galaxies with $10.8 < \log \mathrm{M_*/M_\odot} < 11.2$ (shown with gray circles in the diagram) by the linear regression form using the Bisector method. The best-fitting line is derived as:
\begin{eqnarray}
\log\,({\it SFR}/\mathrm{M_\odot\,yr^{-1}}) &=& \nonumber \\
(1.88\pm0.32) \times \log\,(&M_\mathrm{H_2}&/\mathrm{M_\odot}) - (17.9\pm3.1),
\label{eq:Mh2_SFR}
\end{eqnarray}
which is indicated by the broken line in the left panel of Figure~\ref{Mh2_SFR}.
We then measure the distance to this best-fit line in the orthogonal direction for each galaxy, and perform the two-sample KS test between the disk+green and bulge+green galaxies to test if there is any difference in zero points and/or dispersion between the correlations for disk+green and bulge+green galaxies. 
The cumulative distribution functions (CDFs) of the distance to the best-fit line are shown in the right panel of Figure~\ref{Mh2_SFR}.
The derived {\it p}-value is 0.50, indicating that it is unlikely that the CO-detected disk+green and bulge+green galaxies are drawn from different parent populations.

Finally, we test the effect of the CO-undetected galaxies in our sample on our results. For this purpose, we include the CO-undetected galaxies into the KS test assuming two extreme cases.
The first approach is to use the upper limit values of $f_\mathrm{H_2}$ for CO-undetected galaxies, and in this case we obtain $p = 0.54$ from the KS test. The second approach is to assume their $f_\mathrm{H_2}$ to be 1\,dex smaller than the minimum $f_\mathrm{H_2}$ for the CO-detected galaxies ($\sim 0.1\,\%$), and we derive $p=0.25$ in this case. In these ways, we suggest that the result is unchanged even if we include the CO-undetected galaxies into the KS test.
We therefore conclude that the majority of green-valley galaxies (at fixed stellar mass and SFR) have the same SFE, independent of their galaxy morphologies. 
We note that some fraction of the bulge-dominated galaxies have different properties from other galaxies (as represented by the CO-undetected sources identified in our bulge+green galaxies), and they are an interesting population against the morphological quenching scenario, although the number of such exceptional sources is small ($\sim 23 \%$) at this moment.

%%%%%%%%%%  END: RESULTS  %%%%%%%%%%

%%%%%%%%%%  START: DISCUSSION  %%%%%%%%%%
\section{Discussion} \label{sec:discussion}

\subsection{Morphological Dependence of SFE for Main Sequence Galaxies} \label{subsec:MS}

We tried to test the real effect of galaxy morphologies on SF quenching, by comparing the SFE of green-valley galaxies {\it at fixed $M_*$ and sSFR} with distinct morphologies of ``disk+green" and ``bulge+green" galaxies. Our observations demonstrate that the SFE distribution of green-valley galaxies does {\it not} significantly change with their galaxy morphologies (although there exist a few outliers showing higher SFE in the bulge+green sample). In other words, we do not see the systematic decrease of SFE in bulge-dominated galaxies as reported by previous studies, at least when we study green-valley galaxies at fixed SFR and $M_*$.

An interesting examination to be performed as a next step is to study whether or not our results are universal for galaxies with different SF levels (e.g.\ main-sequence (MS) galaxies).
To this end, we use all the CO data of the star-forming main-sequence (MS) galaxies in the xCOLD~GASS sample located within $\pm$0.6~dex (corresponding to $2\sigma$) from the local MS relation defined by \citet{Elb07}:
\begin{equation}
\log ({\it SFR}/\mathrm{M_\odot\,yr^{-1}}) = 0.77 \times \log (M_*/\mathrm{M_\odot}) - 7.53.
\end{equation}

\begin{figure}
\plotone{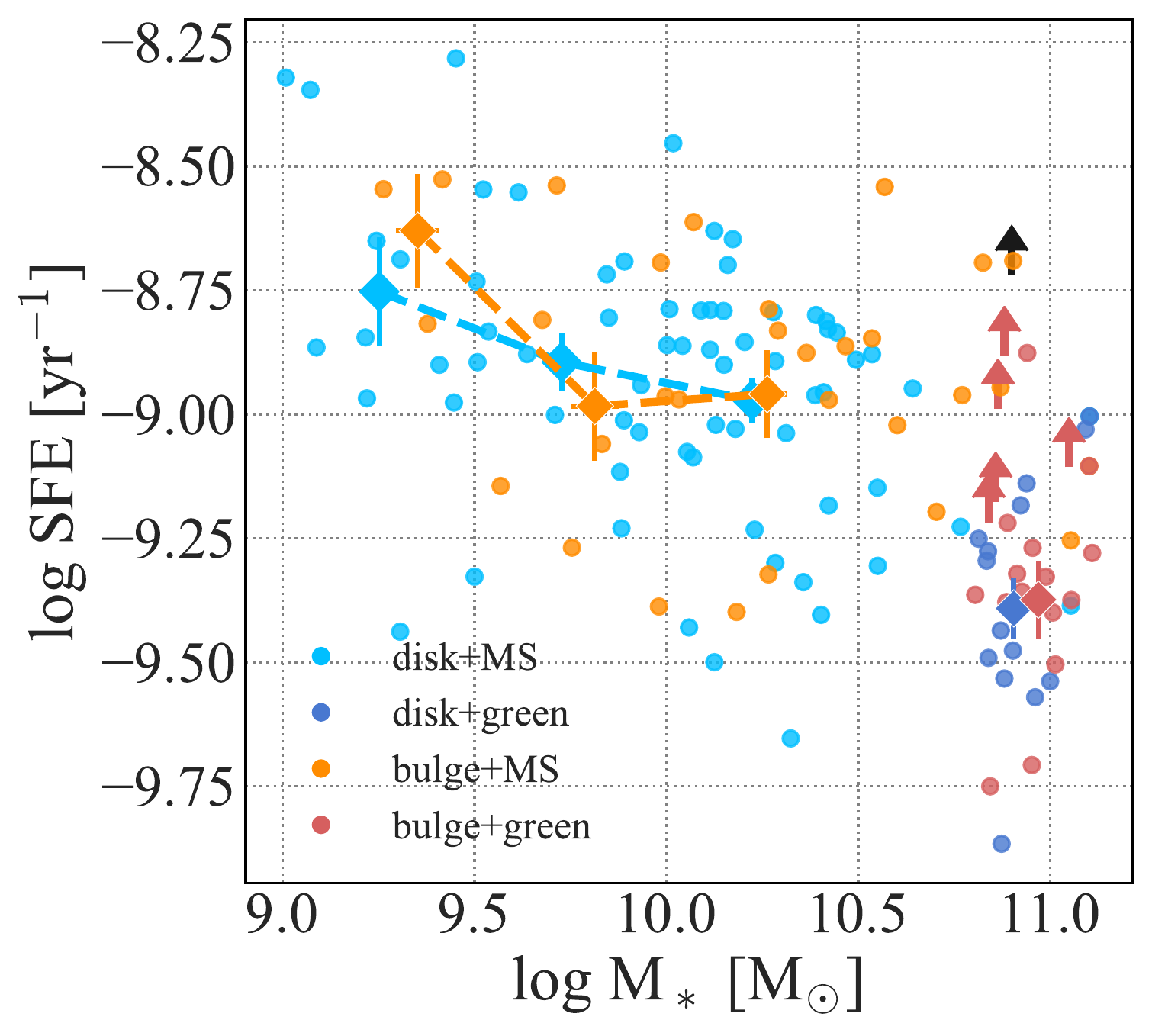}
\caption{Distribution of galaxies on the M$_*$--SFE plane for the disk-dominated MS (disk+MS, cyan), disk-dominated green-valley (disk+green, blue), bulge-dominated MS (bulge+MS, orange), and bulge-dominated green-valley galaxies (bulge+green, red). Diamonds and their error bars show the mean and standard deviation for M$_*$ and SFE, where the MS galaxies are divided into three M$_*$ bins ($\log (M_*/\mathrm{M_\odot})= 9.0-9.5$, $9.5-10.0$ and $10.0-10.5$). As for Figure~\ref{Mh2_SFR}-left, the red and black arrows show the lower limits of SFE for CO-undetected bulge+green galaxies and their stacking result, respectively.}
\label{summary}
\end{figure}

In Figure~\ref{summary}, we plot all the galaxies including the MS and green-valley galaxies with bulge- and disk-dominated morphologies on the $M_*$--SFE plane.
Here, the MS galaxies are classified as bulge-dominated with C-index $>2.8$ and disk-dominated with C-index $<2.2$ as with the green-valley galaxies.
We also perform the KS test of SFE distributions for bulge- and disk-dominated galaxies on the MS galaxies, and derive $p = 0.96$, demonstrating that there is no morphological difference in SFE for the MS galaxies.
This is consistent with our results on the green-valley galaxies, and further in agreement with \citet{Sai17} who report that the molecular gas depletion timescale for galaxies on and above the MS does not strongly depend on stellar mass surface density. Although \citet{Sai17} showed a positive correlation between stellar mass surface density and molecular gas depletion timescale by using all the xCOLD~GASS galaxies, our results suggest that the trend may be primarily driven by the general trend between the morphologies and the evolutionary stages of galaxies, as we pointed out in Secion~\ref{sec:intro}.
It is also evident from Figure~\ref{summary} that there is a significant drop of the average SFEs from the MS to the green-valley phase for both bulge- and disk-dominated galaxies. 
Our results suggest that physical mechanisms triggering SF quenching equally decrease the SFE for both disk- and bulge-dominated galaxies.

\subsection{Interpretations of CO-undetected Sources} \label{subsec:outliers}

As shown in the previous sections, we identified some outliers showing exceptionally high SFEs in our bulge+green sample. In this section, we try to discuss the nature and origin of these outliers.
We first investigate the effects of uncertainties in SFR.
We have used the SFRs computed by the MPA/JHU group mainly based on the optical emission lines \citep{Bri04}.
However, they might have large uncertainties in SFR because their measurements are primarily based on the spectra observed with SDSS 3$\arcsec$ fiber.
In fact, typical galaxy sizes of our sample are $\sim$15$\arcsec$ (see Figure~\ref{spectra}), meaning that the aperture correction is very large, which may affect the robustness of our results.
Furthermore, SFRs of galaxies without strong emission lines (hence low SFRs) are derived assuming a stellar-mass dependent dust attenuation obtained from all the SDSS star-forming galaxies.
It is expected that SFRs derived with this approach would also have large uncertainties.
In fact, some of our target green-valley galaxies do not show significant optical emission lines to properly estimate their dust attenuation level.
It is therefore important to verify our results by using some independent SFR measurements.

We attempt to use the infrared luminosity ($L_\mathrm{IR}$) as an alternative SFR tracer.
The $L_\mathrm{IR}$ is estimated from the \textit{Wide-field Infrared Survey Explorer} \citep[\textit{WISE;}][]{Wri10} 22\,$\mu$m (W4 band) photometry provided by \citet{Sal16}, where they derive $L_\mathrm{IR}$ of individual galaxies by fitting the WISE 22\,$\mu$m (W4 band) luminosities to the luminosity-dependent IR templates of \citet{Cha01} (see \citet{Sal16} for more details on the methodology and robustness of their $L_\mathrm{IR}$ measurements).
We note that we use the W4 magnitude measured in the elliptical aperture (\texttt{w4gmag}) in the ALLWISE source catalog for extended sources with ${\tt etx\_flg} > 0$\footnote{Note that ${\tt etx\_flg}$ is a flag of extended sources, where ${\tt etx\_flg}$$>0$ indicates an extended source.}, while we use the profile-fitting photometry (\texttt{w4mpro}) for galaxies with ${\tt etx\_flg}=0$.
We then derive the IR-based SFRs by using the conversion equation established by \citet{Ken98} adjusted to the \citet{Kro01} IMF:
\begin{equation}
\log\,({\it SFR}/\mathrm{M_\odot\,yr^{-1}}) = \log\,(L_{IR}/\mathrm{L_\odot}) - 9.91.
\end{equation}
The SFRs derived from L$_\mathrm{IR}$ (SFR$_{\it WISE}$) for the individual sources are also listed in Table \ref{SP}.

\begin{figure*}
\plottwo{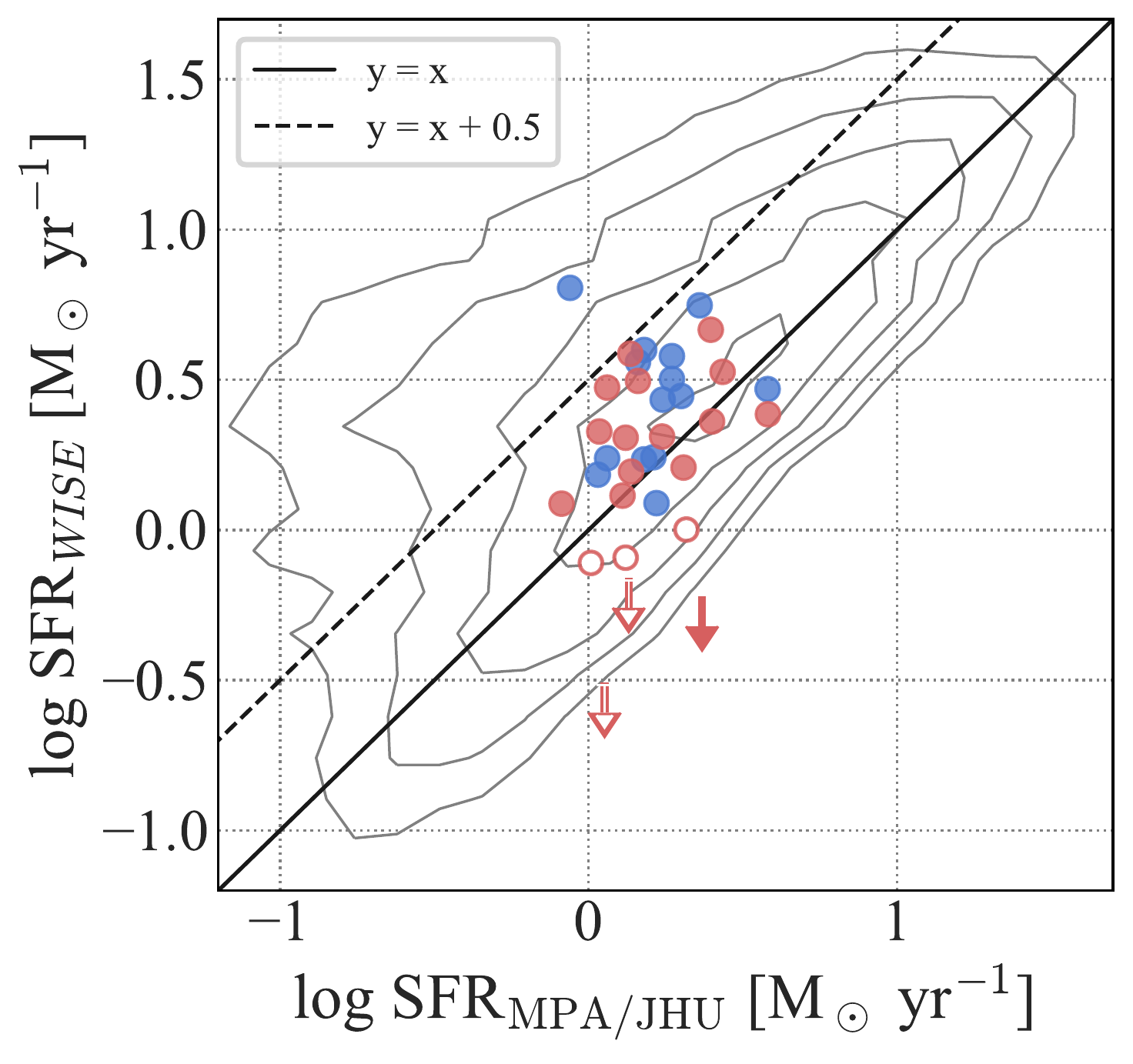}{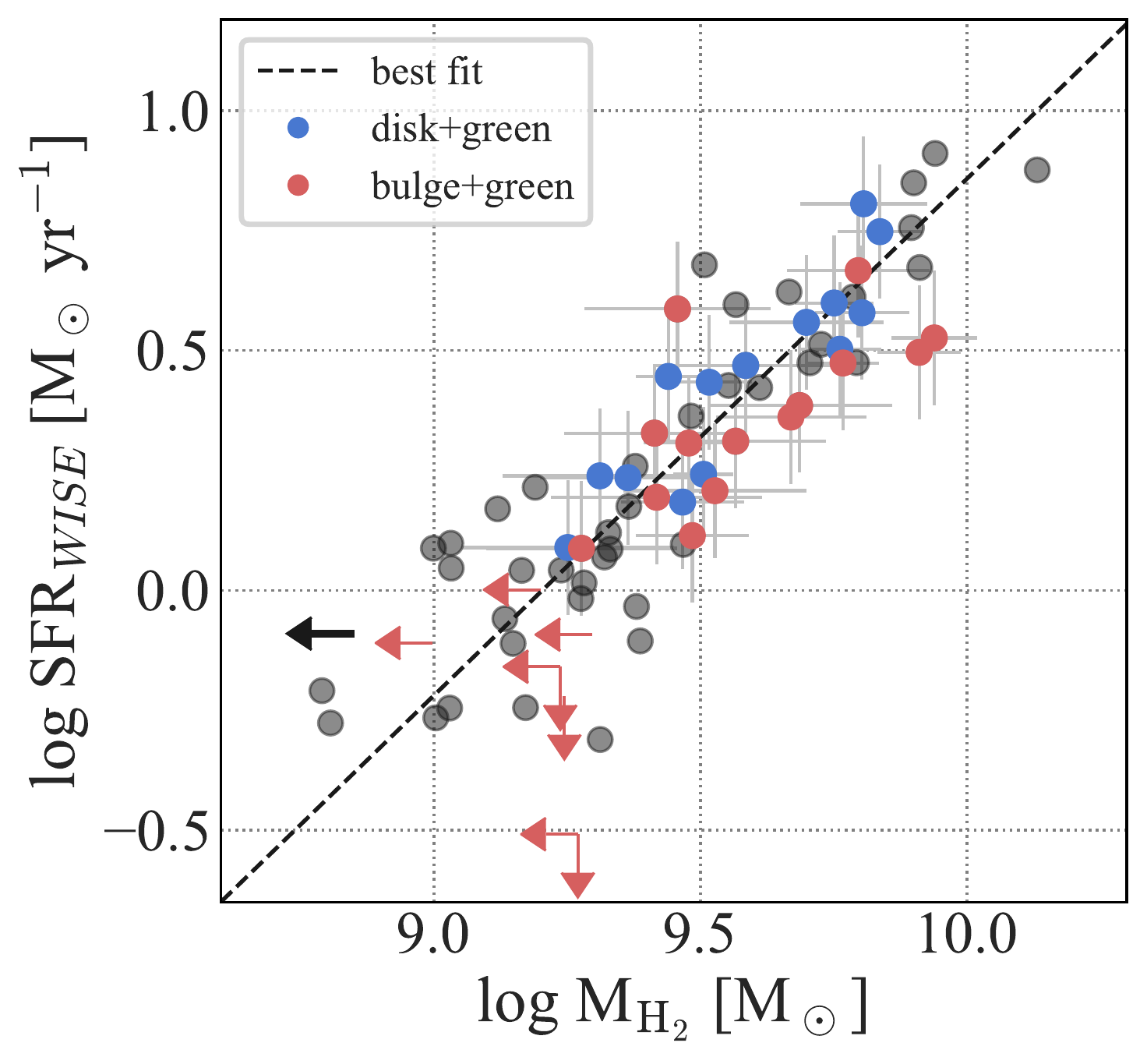}
\caption{(Left): Comparison between the SFRs from the MPA/JHU catalog and those from \textit{WISE} 22\,$\mu$m photometry for our sample. The red and blue symbols indicate the bulge+green and disk+green galaxies, respectively. The circles and arrows show the SFR$_{\it WISE}$ for 22\,$\mu$m detected sources and the 3$\sigma$ upper limits of SFR$_{\it WISE}$ for 22\,$\mu$m undetected sources, respectively. The open symbols indicate the CO-undetected sources.
(Right): The relation between M$_\mathrm{H_2}$ and the IR-based SFRs derived with WISE 22 $\mu$m photometry.
The meanings of the symbols are the same as in Figure~\ref{Mh2_SFR}. 
The black arrow shows the result of our stacking analysis for the CO-undetected sources. We note that we derive their average SFR$_{\it WISE}$ by creating a stacked {\it WISE} W4 image of these CO-undetected sources.
}
\label{SFR_SFR}
\end{figure*}

In the left of Figure~\ref{SFR_SFR}, we compare the SFRs from the MPA/JHU catalog and those from \textit{WISE} data, where the gray contours show the distribution of all SDSS galaxies. The circles and arrows show our samples detected and undetected in {\it WISE}, where the open symbols indicate the CO-undetected galaxies.
It can be seen that the two SFRs are correlated with each other (see the contours in Figure~\ref{SFR_SFR}-left), but this plot demonstrates that SFR$_{\it WISE}$ tend to be higher than SFR$_\mathrm{MPA/JHU}$ (typically by $\sim$0.2--0.3\,dex).
Indeed, most of our samples also show SFR$_{\it WISE}$$>$SFR$_\mathrm{MPA/JHU}$ as shown in the same figure. 
In the right of Figure~\ref{SFR_SFR}, we plot SFR$_{\it WISE}$ against $M_\mathrm{H_2}$. This is the same plot as Figure~\ref{Mh2_SFR}-left, but here we use SFR$_{\it WISE}$ instead of SFR$_\mathrm{MPA/JHU}$. We fitted the data points in the same way as we did in Figure~\ref{Mh2_SFR}, and obtained the best-fit relation of: 
\begin{eqnarray}
\log\,({\it SFR}/\mathrm{M_\odot\,yr^{-1}}) &=& \nonumber \\
(1.08\pm0.09) \times \log\,(&M_\mathrm{H_2}&/\mathrm{M_\odot}) - (9.91\pm0.84).
\label{eq:Mh2_SFRw}
\end{eqnarray}
We also perform the KS test following the procedure of Section~\ref{subsec:Mh2_SFR} and obtain $p=0.04$.
The {\it p}-value is smaller than that derived in the case we use SFR$_\mathrm{MPA/JHU}$, and it is possible that disk+green and bulge+green samples are drawn from different parent population.
However, our result still suggests that, even if two samples are different ($<$0.1-dex), it is significantly smaller than the SFE offset values between late- and early-type galaxies (a factor of $\sim 2$; i.e. $\sim 0.3$~dex) reported by previous studies \citep[][]{Mar13, Dav14} as mentioned in Section~\ref{sec:intro}.
We also compare the SFE distributions, re-defined by SFR$_{\it WISE}$, between disk+green and bulge+green samples, and confirm that there is no significant difference ($p \sim 0.7$).
After all, an important message from the result of this analysis is that our conclusion on the morphological independence of SFEs of the green-valley galaxies are not strongly affected by the choice of SFR tracers.

In the left panel of Figure~\ref{SFR_SFR}, we find that the CO-undetected galaxies (open symbols) have lower SFR$_{\it WISE}$ than SFR$_\mathrm{MPA/JHU}$. These CO-undetected sources are shown with the red leftward arrows in the right panel of Figure~\ref{SFR_SFR} (upper limits for their M$_\mathrm{H_2}$), and in this case most of the CO-undetected galaxies do not contradict with the best-fit line of the $M_\mathrm{H_2}$--SFR$_{\it WISE}$ relation. We caution that those CO-undetected sources might be simply low SFR galaxies, rather than the molecular gas deficient sources as we claimed in the previous sections. Nevertheless, we stress that the result from our stacking analysis for the CO-undetected sources (shown with the black leftward arrow in the right of Figure~\ref{SFR_SFR}) shows a clear offset from the best-fit line, meaning a significantly higher SFE.
Therefore, we still suggest that these CO-undetected galaxies in our bulge+green sample have significantly higher SFEs on average.
They may represent the galaxy population during the SF quenching process due to the presence of bulge component.  

The reason of their molecular gas deficiencies (and unusually high SFEs) is unclear with the available dataset, but we speculate that the fraction of dense molecular gas, which is known to be more tightly correlated with star formation than the gas traced by CO line \citep[e.g.][]{Gao04}, could be increased in these CO-undetected bulge+green galaxies. Observations of such dense cold molecular gas (e.g.\ HCN) of the green-valley galaxies would be our important future work. The molecular gas deficiency may also be explained if the galaxies lose their molecular gas in a shorter timescale than that of their star formation.
However, the change of the dense gas fraction in the bulge+green sample would be a more realistic scenario because it is difficult to decrease the bulk of their molecular gas in a very short time scale considering the typical time scale of star formation ($10^{7-8}$~yr) \citep[e.g.][]{Egu09}. 
Our results suggest that at least a non-negligible fraction ($\sim$20\%) of bulge-dominated green-valley galaxies show a sign of molecular gas deficiency (hence unusually high SFE), but it is clear that we need a larger sample to more quantitatively understand the fraction and the nature of such interesting population, and to firmly conclude that this population appears only in bulge-dominated galaxies.

\subsection{Comparison with Previous Studies} \label{subsec:comparison}

The morphological quenching scenario, proposed by \citet{Mar09}, has been supported by many recent studies with CO observations.
For example, \citet{Sai11b} reported that the gas depletion time $\tau_{dep} (= M_\mathrm{H_2}/{\it SFR})$ increases with C-index of galaxies.
\citet{Mar13, Dav14} also investigated the relationship between $\Sigma_\mathrm{HI+H_2}$ and $\Sigma_\mathrm{SFR}$ for local early-type galaxies, and suggested that star formation in early-type galaxies are less efficient by a factor of $\sim$2 than those in late-type galaxies.
More recently, \citet{Col18} showed that the spatially resolved gas depletion time $\tau^{mol}_{dep} (= \Sigma_\mathrm{gas}/\Sigma_\mathrm{SFR})$ of elliptical galaxies tends to be longer than that of galaxies with spiral galaxies.
However, it should be noted that those earlier works supporting the morphological quenching scenario attempted to compare typical early-type and late-type galaxies. 
``Early-type'' galaxies are usually selected from passive galaxy population on the red sequence, while ``late-type'' galaxies are selected from star-forming population in the blue cloud (as mentioned in Section~\ref{sec:intro}).

In this paper, we reported that there is no significant difference in the SFE for bulge- and disk-dominated green-valley galaxies (as well as MS galaxies). We note, however, that our results do not necessarily suggest that galaxy morphologies do not affect the SF quenching process at all. As mentioned in Section~\ref{subsec:outliers}, we find some CO-undetected galaxies with unusually high SFEs in our bulge+green sample, which can be identified as outliers in the M$_\mathrm{H_2}$-SFR diagram. Those molecular gas deficient galaxies with bulge-dominated morphologies may represent the galaxy population during the SF quenching process due to the presence of bulge component.  

\citet{Sch14} suggested that the green-valley galaxies with bulge-dominated morphologies are quenched with shorter time scale than those with late-type morphologies.
Interestingly, \citet{Fab11}, who studied the morphological dependence of atomic gas contents in galaxies, claimed that early-type galaxies tend to have a smaller amount of atomic gas than the late-type galaxies at fixed M$_*$ and SFR, and it is in contradiction with morphological quenching scenario.
As mentioned in \citet{Mar13}, the atomic gas component is distributed in the extended disks of galaxies, and it would not be an ideal tool to discuss the SF quenching process.
Nevertheless, our CO observations also revealed that there is no significant difference in the average SFE for the bulge- and disk-dominated galaxies at fixed M$_*$ and SFR,
and that there exist 20~\% of the bulge+green galaxies with exceptionally higher SFE and lower molecular gas fraction than the other bulge+green galaxies, consistent with the trend reported by \citet{Fab11}.
Overall, we suggest that the presence of bulge component in galaxies should affect the molecular gas, in the sense that galaxies with a prominent bulge tend to rapidly remove and/or consume their molecular gas content (hence showing apparently higher SFEs), in contrast to the prediction of the morphological quenching scenario.

Finally, we comment that our results are based on the {\it spatially integrated} stellar mass, molecular gas mass, and SFR information, while it is very important to perform a systematic study of {\it spatially resolved} molecular gas properties in green-valley galaxies using interferometric data (e.g.\ ALMA or NOEMA), in order to directly assess the morphological impacts on the individual star-forming regions within the galaxies. \citet{Lin17} recently performed a pioneering work of the spatially resolved star formation and molecular gas properties of three local green-valley galaxies with ALMA and MaNGA data, and they reported that both $f_\mathrm{H_2}$ and SFE play a role in lowering the sSFR in the disk component of green-valley galaxies.
It is important to extend this survey to galaxies with various morphological types to reveal the real morphological impacts on SF quenching process inside the galaxies.

%%%%%%%%%%  END: DISCUSSIONS  %%%%%%%%%%

%%%%%%%%%%  START: CONCLUSION  %%%%%%%%%%
\section{Conclusion} \label{sec:conclusion}

We present our new CO observations of the green-valley galaxies with different morphologies (``disk+green'' and ``bulge+green'' samples) at fixed M$_*$ and SFR using the NRO~45m telescope, to investigate the real effect of galaxy morphologies on their star formation efficiency as well as the correlation between M$_\mathrm{H_2}$ and SFR. Our findings are summarized as follows.

\begin{enumerate}
\item The distribution of M$_\mathrm{H_2}$, $f_\mathrm{H_2}$ and SFE for the CO-detected green-valley galaxies ($10.8<\log\,(M_*/\mathrm{M_\odot})<11.2$ and $-11< \log\,(sSFR/\mathrm{yr^{-1}}) <-10.5$) are independent of their morphologies.
We also found that both CO-detected disk-dominated and bulge-dominated green-valley galaxies follow the identical correlation between $M_\mathrm{H_2}$ and SFR.
There is no evidence for the decline in the SFE of galaxies with early-type morphologies. 

\item We did not detect CO line from $\sim$20\% (5 out of 22) of the bulge+green galaxies. Even in the stacked spectrum of CO-undetected galaxies, the emission line was not detected.
These CO-undetected galaxies show exceptionally {\it high} SFEs for their M$_*$ and SFR, in contrast to the prediction from the morphological quenching scenario.
Although we confirmed that our main results on the morphological independence of the averaged SFEs of galaxies (at fixed M$_*$ and SFR) are not changed by these CO-undetected sources, those molecular gas deficient sources identified only in the bulge+green sample might be an important population during the phase of SF quenching due to the presence of stellar bulge in galaxies. 

\item Using the CO data from the xCOLD~GASS survey, we found that our results on the morphological independence of SFE is also valid for normal star-forming galaxy population on the main sequence; i.e.\ at a fixed M$_*$, the average SFE of galaxies on the star-forming main sequence does not vary with their morphologies. This is in agreement with \citet{Sai17} who used the stellar mass surface density as a morphological indicator, rather than the C-index.
We confirmed that there is a significant decrease of the mean SFEs from the main sequence to the green-valley phase, but the level of this decline is the same for both bulge- and disk-dominated galaxies. This result suggests that the SF quenching mechanism equally affects the SFE of galaxies, regardless of their morphological properties.

\end{enumerate}

We emphasize again that our target galaxies are carefully selected from a small window on the SFR--M$_*$ diagram, whilst many other studies discussing the morphological difference in SFE compare early-type galaxies selected from passive (red-sequence) galaxy population and late-type galaxies selected from star-forming (blue-cloud) galaxies.  
Although there still remains a possibility that the SF quenching time scale is different between the disk- and bulge-dominated galaxies, our results suggest that the galaxy morphology is not a primary factor to control the SFE of galaxies as long as we focus on galaxies at fixed M$_*$ and SFR.

%%%%%%%%%%  END: CONCLUSION  %%%%%%%%%%

\acknowledgments
We thank the anonymous referee for careful reading and useful comments that helped to improve our paper.

This research is based on SDSS-III, and observations at the Nobeyama 45-m radio telescope.

Funding for SDSS-III has been provided by the Alfred P. Sloan Foundation, the Participating Institutions, the National Science Foundation, and the U.S. Department of Energy Office of Science. The SDSS-III web site is \url{http://www.sdss3.org/}.
SDSS-III is managed by the Astrophysical Research Consortium for the Participating Institutions of the SDSS-III Collaboration including the University of Arizona, the Brazilian Participation Group, Brookhaven National Laboratory, University of Cambridge, Carnegie Mellon University, University of Florida, the French Participation Group, the German Participation Group, Harvard University, the Instituto de Astrofisica de Canarias, the Michigan State/Notre Dame/JINA Participation Group, Johns Hopkins University, Lawrence Berkeley National Laboratory, Max Planck Institute for Astrophysics, Max Planck Institute for Extraterrestrial Physics, New Mexico State University, New York University, Ohio State University, Pennsylvania State University, University of Portsmouth, Princeton University, the Spanish Participation Group, University of Tokyo, University of Utah, Vanderbilt University, University of Virginia, University of Washington, and Yale University.

The Nobeyama 45-m radio telescope is operated by Nobeyama Radio Observatory, a branch of National Astronomical Observatory of Japan.

This work was financially supported in part by a Grant-in-Aid for the Scientific Research (No.\,26800107; 18K13588) by the Japanese Ministry of Education, Culture, Sports and Science.

\facilities{NRO\,45m}
\software{NEWSTAR}

%%%%%%%%%%  START: REFERENCES  %%%%%%%%%%

%%%%%%%%%%  END: REFERENCES  %%%%%%%%%%

\end{document}